\documentclass[fleqn,usenatbib]{mnras}

\usepackage{newtxtext,newtxmath}

\usepackage[T1]{fontenc}

\DeclareRobustCommand{\VAN}[3]{#2}
\let\VANthebibliography\thebibliography
\def\thebibliography{\DeclareRobustCommand{\VAN}[3]{##3}\VANthebibliography}

\usepackage{graphicx}	
\usepackage{adjustbox}  
\usepackage{amsmath}	
\usepackage{wasysym}    
\usepackage{bm}         
\usepackage{xspace}
\usepackage{xcolor}
\makeatletter 
  \patchcmd{\NAT@citex}
    {\@citea\NAT@hyper@{%
      \NAT@nmfmt{\NAT@nm}%
      \hyper@natlinkbreak{\NAT@aysep\NAT@spacechar}{\@citeb\@extra@b@citeb}%
      \NAT@date}}
    {\@citea\NAT@nmfmt{\NAT@nm}%
    \NAT@aysep\NAT@spacechar\NAT@hyper@{\NAT@date}}{}{}

  \patchcmd{\NAT@citex}
    {\@citea\NAT@hyper@{%
      \NAT@nmfmt{\NAT@nm}%
      \hyper@natlinkbreak{\NAT@spacechar\NAT@@open\if*#1*\else#1\NAT@spacechar\fi}%
        {\@citeb\@extra@b@citeb}%
      \NAT@date}}
    {\@citea\NAT@nmfmt{\NAT@nm}%
    \NAT@spacechar\NAT@@open\if*#1*\else#1\NAT@spacechar\fi\NAT@hyper@{\NAT@date}}
    {}{}
\makeatother

\newcommand\Msun{\text{M}_{\astrosun}} 
\newcommand\HI{\ion{H}{I}\xspace} 
\newcommand\HII{\ion{H}{II}\xspace} 
\newcommand\thesan{\mbox{\textsc{thesan}}\xspace}
\newcommand\thesanone{\mbox{\textsc{thesan-1}}\xspace}
\newcommand\thesantwo{\mbox{\textsc{thesan-2}}\xspace}
\newcommand\thesanwc{\mbox{\textsc{thesan-wc-2}}\xspace}
\newcommand\thesanhigh{\mbox{\textsc{thesan-high-2}}\xspace}
\newcommand\thesanlow{\mbox{\textsc{thesan-low-2}}\xspace}
\newcommand\thesansdao{\mbox{\textsc{thesan-sdao-2}}\xspace}

\newcommand\orcid[1]{\href{http://orcid.org/#1}{\adjustbox{trim={-.15\width} {0\height} {-.15\width} {0\height},clip}{\includegraphics[height=12pt]{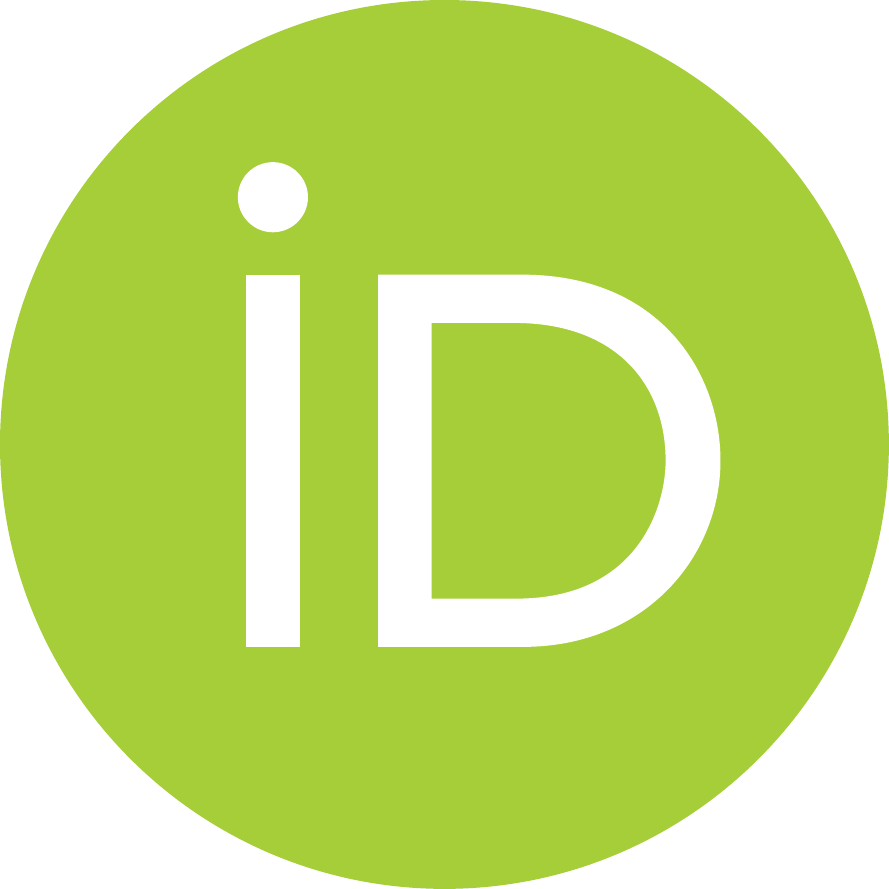}}}}

\title[Ly$\alpha$ luminosity functions]{The \thesan project: Lyman-$\bmath\alpha$ emitter luminosity function calibration}

\author[C.~Xu et al.]{%
Clara~Xu\orcid{0000-0003-4688-6487},$^{1}$\thanks{E-mail: \href{mailto:claraxu@mit.edu}{claraxu@mit.edu}}
Aaron~Smith\orcid{0000-0002-2838-9033},$^{2,1}$\thanks{E-mail: \href{mailto:aaron.smith@cfa.harvard.edu}{aaron.smith@cfa.harvard.edu}; NHFP Einstein Fellow.}
Josh~Borrow\orcid{0000-0002-1327-1921},$^{1}$
Enrico~Garaldi\orcid{0000-0002-6021-7020},$^{3}$
Rahul~Kannan\orcid{0000-0001-6092-2187},$^{2,4}$
\newauthor%
Mark~Vogelsberger\orcid{0000-0001-8593-7692},$^{1,5}$
R\"{u}diger~Pakmor\orcid{0000-0003-3308-2420},$^{3}$
Volker~Springel\orcid{0000-0001-5976-4599}$^{3}$
and Lars~Hernquist$^{2}$
\\%
\\%
$^{1}$Department of Physics, Massachusetts Institute of Technology, Cambridge, MA 02139, USA \\%
$^{2}$Center for Astrophysics $\vert$ Harvard $\&$ Smithsonian, 60 Garden Street, Cambridge, MA 02138, USA \\%
$^{3}$Max-Planck Institute for Astrophysics, Karl-Schwarzschild-Str.~1, D-85741 Garching, Germany \\%
$^{4}$Department of Physics and Astronomy, York University, 4700 Keele St., Toronto, Ontario, Canada, MJ3 1P3 \\%
$^{5}$The NSF AI Institute for Artificial Intelligence and Fundamental Interactions, Massachusetts Institute of Technology, Cambridge MA 02139, USA
}

\date{Accepted 2023 March 11. Received 2023 March 08; in original form 2022 November 03}

\pubyear{2023}

\begin{document}
\label{firstpage}
\pagerange{\pageref{firstpage}--\pageref{lastpage}}
\maketitle

\begin{abstract}
The observability of Lyman-alpha emitting galaxies (LAEs) during the Epoch of Reionization can provide a sensitive probe of the evolving neutral hydrogen gas distribution, thus setting valuable constraints to distinguish different reionization models. In this study, we utilize the new \textsc{thesan} suite of large-volume ($L_\text{box} = 95.5\,\text{cMpc}$) cosmological radiation-hydrodynamic simulations to directly model the Ly$\alpha$ emission from individual galaxies and the subsequent transmission through the intergalactic medium. \textsc{thesan} combines the \textsc{arepo-rt} radiation-hydrodynamic solver with the IllustrisTNG galaxy formation model and includes high- and medium-resolution simulations designed to investigate the impacts of halo-mass-dependent escape fractions, alternative dark matter models, and numerical convergence. We find important differences in the Ly$\alpha$ transmission based on reionization history, bubble morphology, frequency offset from line centre, and galaxy brightness. For a given global neutral fraction, Ly$\alpha$ transmission reduces when low mass haloes dominate reionization over high mass haloes. Furthermore, the variation across sightlines for a single galaxy is greater than the variation across all galaxies. This collectively affects the visibility of LAEs, directly impacting observed Ly$\alpha$ luminosity functions (LFs). We employ Gaussian Process Regression using \textsc{SWIFTEmulator} to rapidly constrain an empirical model for dust escape fractions and emergent spectral line profiles to match observed LFs. We find that dust strongly impacts the Ly$\alpha$ transmission and covering fractions of $M_\text{UV} \lesssim -19$ galaxies in $M_\text{vir} \gtrsim 10^{11}\,\Msun$ haloes, such that the dominant mode of removing Ly$\alpha$ photons in non-LAEs changes from low IGM transmission to high dust absorption around $z \sim 7$.
\end{abstract}




\section{Introduction}
The prominent Lyman-alpha~(Ly$\alpha$) emission line of atomic hydrogen can serve as a powerful probe of the high-redshift Universe \citep{Partridge1967,Ouchi2020}. The intrinsic Ly$\alpha$ luminosity closely traces the global star-formation rate (SFR) of galaxies on relatively short timescales ($\lesssim$\,10\,Myr) throughout cosmic history \citep{KennicuttEvans2012,Dayal2018}. These Ly$\alpha$ photons then undergo a complex resonant scattering process coupling the emission to the neutral hydrogen and dust column densities, kinematics, and geometry, such that the emergent spectra and escape fractions can be difficult to model and interpret \citep{Dijkstra2019}. In addition, neutral hydrogen in the intergalactic medium (IGM) far from the source can remove photons in the vicinity of the Ly$\alpha$ line with a single scattering out of the line of sight \citep{GunnPeterson1965}. These effects are particularly strong during the Epoch of Reionization (EoR), before radiation from stars and galaxies completely ionized the hydrogen throughout the Universe \citep{Barkana2001,Wise2019}. At these high redshifts Ly$\alpha$ photons are typically only transmitted through the IGM if they scatter in frequency to the red damping wing of the line profile \citep{Miralda-Escude1998,MadauRees2000,Mesinger2015}. Together these phenomena allow us to leverage Ly$\alpha$ emitting galaxies (LAEs) as a probe of reionization \citep[e.g.][]{MalhotraRhoads2004,McQuinn2007,Dijkstra2014,Kakiichi2016}.

Surveys of high-redshift LAEs have made significant progress over the past decades, overcoming challenging conditions for observing the Ly$\alpha$ Universe \citep[for a review see][]{Ouchi2020}. Improved strategies and technologies have helped to reach greater sensitivities, cover larger areas, or mitigate atmospheric and foreground contamination \citep[e.g.][]{Rhoads2000,Taniguchi2005,Finkelstein2009}. At intermediate redshifts ($z \sim 2$--$6$) these now include novel integral field units (IFUs) to increase the number of known LAEs and better understand their detailed spatial--spectral Ly$\alpha$ characteristics. For example, the MUSE Integral Field Spectrograph on the Very Large Telescope \citep[VLT;][]{Bacon2010,Bacon2017} and Hobby--Eberly Telescope Dark Energy Experiment \citep[HETDEX;][]{Adams2011,Gebhardt2021} among others are providing complementary insights about large-scale structures and circumgalactic environments. At higher redshifts we rely on narrow-band surveys such as SILVERRUSH mapping over 2000 LAEs at $z = 5.7$ and $6.6$ (corresponding to advantageous low-contamination windows), which reveal target halo masses of $M_\text{halo} \sim 10^{11}\,\Msun$ \citep{Ouchi2018}. The Ly$\alpha$ equivalent widths are $\gtrsim20\,\text{\AA}$ for a small subset of the galaxy population (per cent level) as a result of either low duty cycles of bursty star-formation or small covering fractions of advantageous viewing angles. Spectroscopic searches well into the EoR at $z \gtrsim 7$ are impeded by low number statistics and so far mainly indicate that Ly$\alpha$ transmission is aided by very large ionized bubbles around UV bright galaxies \citep{Jung2020,Jung2022,Endsley2021,Endsley2022}. We anticipate that data from the recently launched \textit{James Webb Space Telescope} (\textit{JWST}) will help unlock additional progress in the study of reionization-era LAEs \citep{Katz2019,Smith2019}. Finally, in addition to observing individual galaxies, future line intensity mapping of integrated diffuse light including Ly$\alpha$ and 21\,cm emission will provide further definitive answers to questions about the EoR \citep{Furlanetto2007,Heneka2017,Hutter2017}.

There is strong motivation to better understand Ly$\alpha$ emission and transmission at $z \gtrsim 5$ from the perspectives of both galaxy formation (astrophysics) and reionization (cosmology). Constraints on the neutral fraction of the IGM can be derived from the evolution of LAEs and the number density of Lyman break galaxies \citep{Ota2008,Ono2012,Jensen2013,Pentericci2014,Tilvi2014,Choudhury2015,Mesinger2015,Sarkar2019}, from the angular correlation function of LAEs \citep{SobacchiMesinger2015}, from the detection of Ly$\alpha$ emission in Lyman break galaxies \citep{Mason2018,Mason2019,Hoag2019,Jung2020}, from a combination of Ly$\alpha$ luminosity, clustering and line profile \citep{Ouchi2010,Goto2021}, from the void probability function \citep{Perez2022}, and from the Ly$\alpha$ visibility \citep{Dijkstra2011}. There are also significant efforts to robustly measure and characterize the observed Ly$\alpha$ luminosity function~(LF) at $z \approx 5$--$7$ \citep{Ouchi2008,Ouchi2010,Santos2016,Konno2018,Khusanova2020,Wold2022} also focusing on the bright end \citep{Taylor2020,Taylor2021}. Interpreting the LF requires understanding both Ly$\alpha$ sources and transmission, which are correlated through the local environment and emission properties of galaxies \citep{Mason2018}. However, the number density of LAEs is also connected to the more accessible ultraviolet luminosity function~(UVLF) through the Ly$\alpha$ luminosity probability distribution for Lyman break galaxies, which allows some understanding of the relative evolution of galaxy and IGM effects \citep{DijkstraWyithe2012,GronkeLF2015,Whitler2020,Morales2021}.

Hydrodynamical simulations have proven to be a powerful framework for modeling the strong impact of frequency-dependent IGM transmission on photons in the vicinity of the Ly$\alpha$ line. Previous work has focused on the statistical nature of transmission curves either for post-reionization ($z \lesssim 5$) spectral modulations \citep{Laursen2011,Byrohl2020} or predictions from large-volume reionization boxes \citep{Gronke2021,Park2021,Smith2022}. Previous works have also explored the integrated Ly$\alpha$ transmission in the context of idealized galaxy spectral profiles \citep{Jensen2013,Weinberger2019,Gangolli2021}. These parameterized models can in principle be tuned to yield Ly$\alpha$ luminosities and equivalent widths that match observations after transmission through the IGM. Such studies based on semi-empirical modelling provide important predictions about the EoR and reveal the effective behaviour of these systems. However, they also gloss over specific details of Ly$\alpha$ radiative transfer, including uncertainties related to the emission and escape of radiation through the interstellar medium (ISM) and circumgalactic medium (CGM) of galaxies. Recently, simulations with higher resolution have been used to directly model the effects of dust absorption and predict equivalent widths, thus enabling a more self-consistent analysis of LAE populations \citep[e.g.][]{Garel2021}. Still, the relatively small box size necessary for the high resolution also leads to difficulty in modeling large-scale effects and the bright end of LFs. Therefore, the empirical and direct transport approaches remain complementary.

In this paper we present a follow-up study of \citet{Smith2022} in which we compare Ly$\alpha$ emission and transmission statistics from all runs from the \thesan suite of reionization simulations \citep{Kannan2022,Garaldi2022,Smith2022}. This allows us to provide additional context for reionization constraints including dependence on frequency, reionization history, and UV brightness. More importantly, we also perform an empirical machine learning study to determine what dust escape fractions and spectrally-averaged IGM transmission properties agree with the observed Ly$\alpha$ LFs. This is intended as a viable alternative to performing sub-grid modelling dependent Ly$\alpha$ radiative transfer calculations, thus acting as a bridge between the trusted but disconnected Ly$\alpha$ intrinsic emission and IGM transmission catalogues. The resulting simulation-based observed Ly$\alpha$ galaxy catalogues will also be publicly available as part of the \thesan data release. In the future we plan to apply this framework to study other Ly$\alpha$-centric probes of reionization.

The remainder of the paper is organized as follows. In Section~\ref{sec:methods}, we briefly describe the \thesan simulations employed throughout this paper including the Ly$\alpha$ catalogues. In Section~\ref{sec:transmission}, we present transmission statistics exploring the non-trivial dependence on frequency, redshift, and UV magnitude. In Section~\ref{sec:LF}, we constrain an empirical model to match observed Ly$\alpha$ LFs and present dust and spectra reprocessed observed transmission statistics. Finally, in Section~\ref{sec:conclusions}, we provide a summary and discussion on Ly$\alpha$ science from \thesan.

\section{Methods}
\label{sec:methods}
In this section, we describe the \thesan simulations used in this work and the calculation of Ly$\alpha$ emission and transmission curves.

\subsection{\thesan simulations}
The \thesan project is a suite of large-volume radiation-magneto-hydrodynamical simulations that self-consistently model the reionization process and the resolved properties of the sources responsible for it. The simulations were performed with \textsc{arepo-rt} \citep{Kannan2019}, a radiation-hydrodynamic extension of the moving mesh code \textsc{arepo} \citep{Springel2010,Weinberger2020}, which solves the fluid equations on an unstructured Voronoi mesh that is allowed to move along with the fluid flow for an accurate quasi-Lagrangian treatment of cosmological gas flows. Gravitational forces are calculated using a hybrid Tree-PM approach, splitting the force into a short-range force that is computed using an oct-tree algorithm \citep{Barnes1986} and a long-range force, estimated using a particle mesh approach.

Radiation fields are modelled using a moment based approach that solves the zeroth and first moments of the radiative transfer equation \citep{Rybicki1986}, coupled with the M1 closure relation, that approximates the Eddington tensor based on the local properties of each cell \citep{Levermore1984}. They are coupled to the gas via a non-equilibrium thermo-chemistry module, which self-consistently calculates the ionization states and cooling rates from hydrogen and helium, while also including equilibrium metal cooling and Compton cooling of the CMB \citep[see Section 3.2.1 of][for more details]{Kannan2019}. Both stars and AGN act as sources of radiation, with the spectral energy distribution of stars taken from the Binary Population and Spectral Synthesis models \citep[BPASS version 2.2.1;][]{BPASS2017}, assuming a Chabrier IMF \citep{Chabrier2003}. The AGN radiation output is scaled linearly with the mass accretion rate with a radiation conversion efficiency of 0.2 \citep{Weinberger2018} and a \citet{Lusso2015} parametrization for the shape of its spectrum. For computational efficiency, we choose to only model the ionizing part of the radiation spectrum and discretize photons into energy bins defined by the following thresholds: $[13.6, 24.6, 54.4, \infty)\,\text{eV}$. Each resolution element tracks the comoving photon number density and flux for each bin. Finally, we employ a reduced speed of light approximation with a value of $\tilde{c} = 0.2\,c$ \citep[][]{Gnedin2016}, which is large enough to accurately capture the propagation of ionization fronts and post-reionization gas properties.

The prescriptions for processes happening below the resolution limit of the simulations, such as star and black hole formation and feedback and metal production and enrichment, are taken from the IllustrisTNG model \citep{Weinberger2017,Springel2018,Pillepich2018a,Pillepich2018b,Naiman2018,Marinacci2018,Nelson2018,Nelson2019,Pillepich2019}. The model is augmented with a scalar dust model that tracks the production, growth, and destruction of dust using the formalism outlined in \citet{McKinnon2017}. An additional birth cloud escape fraction parameter, $f_\text{esc}^\text{cloud}=0.37$, is added to mimic the absorption of LyC photons happening below the resolution scale of the simulation. The parameter is tuned such that the simulation reproduces a realistic late-reionization history \citep{Kannan2022}, which matches the observed neutral fraction evolution in the Universe \citep{Greig2017}.

\begin{figure}
  \centering
  \includegraphics[width=\columnwidth]{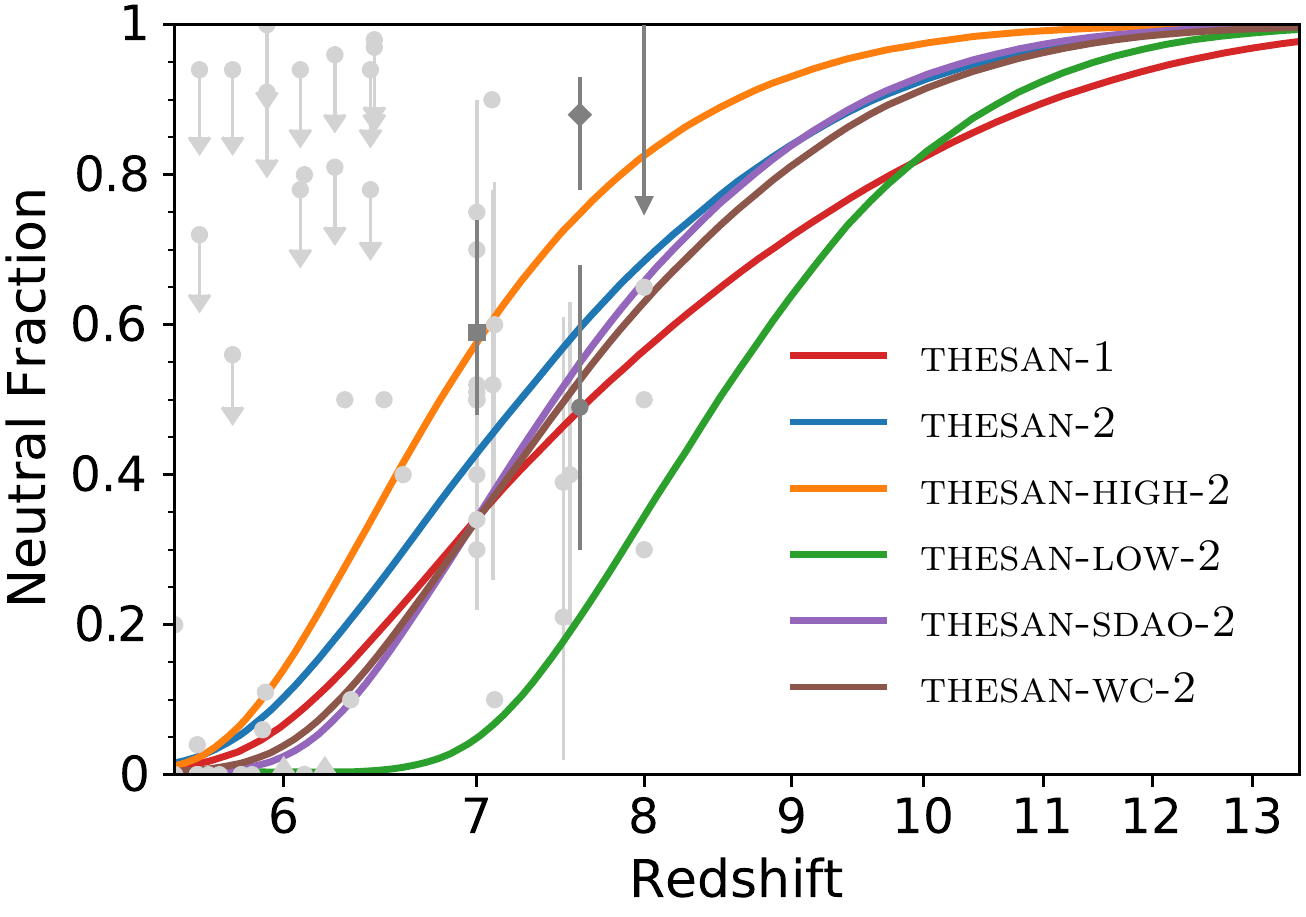}
  \caption{Reionization history of each simulation plotted as neutral fraction against redshift. \thesanhigh and \thesanlow bracket the other simulations, with \thesanlow reionizing earlier than the others due to the greater contribution of low-mass galaxies to reionization, whereas in \thesanhigh the rarer high-mass galaxies dominate. \thesantwo, \thesanwc, and \thesansdao have similar reionization histories, though as expected \thesanwc is closest to \thesanone. For reference, we also include observational constraints from the detection of Ly$\alpha$ emission in Lyman break selected galaxies (\citealp{Mason2018z7} -- square; \citealp{Mason2019} -- triangle; \citealp{Hoag2019} -- diamond; \citealp{Jung2020} -- circle). Constraints on the reionization history from other methods are included in light grey.}
  \label{fig:reion_history}
\end{figure}

\begin{figure*}
  \centering
  \includegraphics[width=\textwidth]{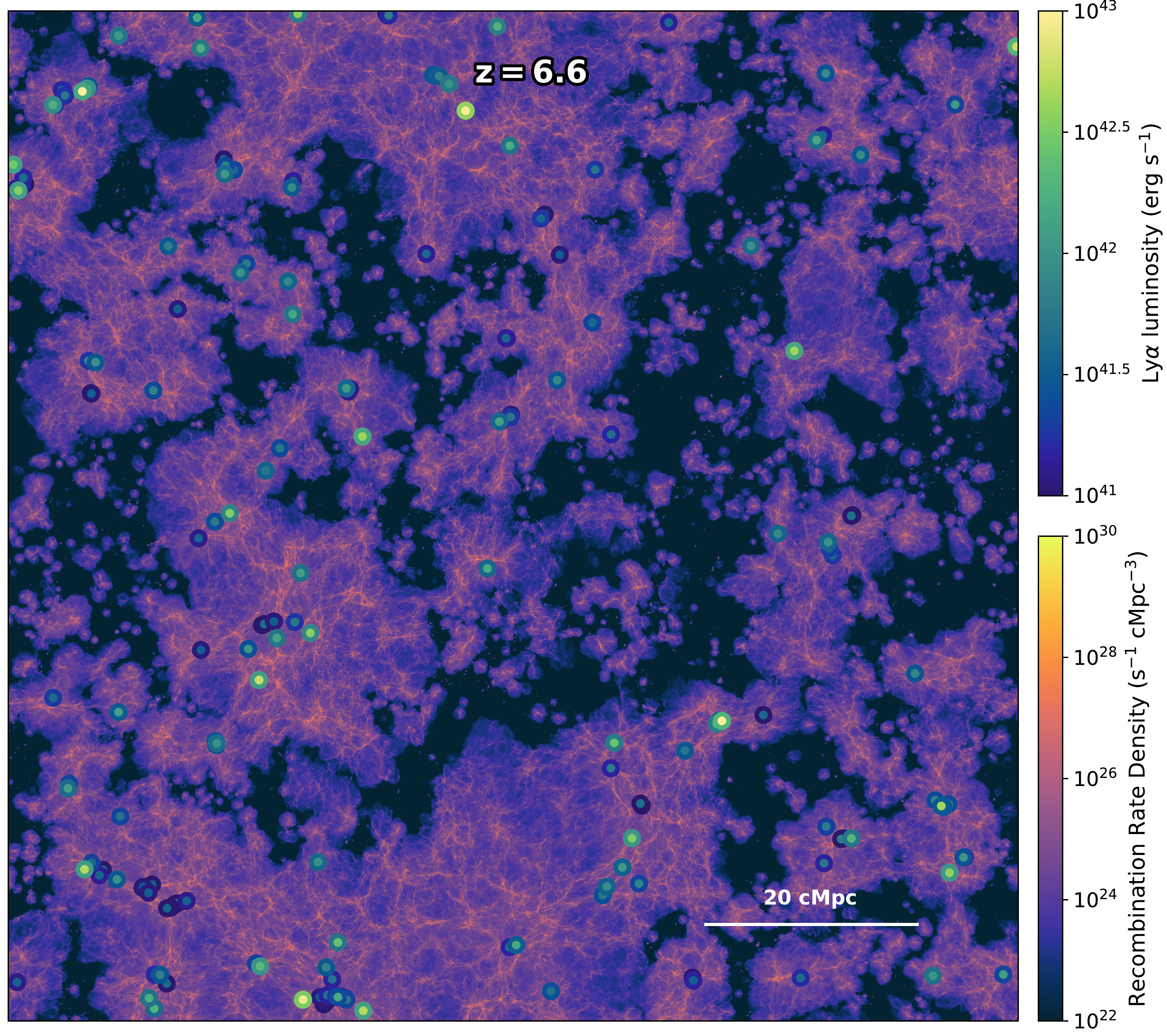}
  \caption{Recombination rate density in \thesanone at $z=6.6$ covering the middle 2.5$\%$ of the box volume. The locations of galaxies with intrinsic Ly$\alpha$ luminosity greater than $10^{41}\,\text{erg\,s}^{-1}$ are plotted over the background, with the centre (edge) colour indicating the intrinsic (observed) luminosity as calculated in section \ref{sec:LF}. Galaxies with high Ly$\alpha$ emission are clearly correlated with the overall large-scale structure of the simulation.}
  \label{fig:visualization}
\end{figure*}

All \thesan simulations follow the evolution of a cubic patch of the universe with linear comoving size \mbox{$L_\text{box} = 95.5$\,cMpc}, and utilize variance-suppressed initial conditions \citep{Angulo2016}. We employ a \citet{Planck2015_cosmo} cosmology (more precisely, the one obtained from their \texttt{TT,TE,EE+lowP+lensing+BAO+JLA+H$_0$} dataset), i.e. $H_0 = 100\,h\,\text{km\,s}^{-1}\text{Mpc}^{-1}$ with $h=0.6774$, $\Omega_\mathrm{m} = 0.3089$, $\Omega_\Lambda = 0.6911$, $\Omega_\mathrm{b} = 0.0486$, $\sigma_8 = 0.8159$, and $n_s = 0.9667$, where all symbols have their usual meanings. The highest resolution \thesanone simulation has a total number of dark matter and (initial) gas particles of $2100^3$ each with mass resolutions of $m_\text{DM} = 3.12 \times 10^6\,\Msun$ and $m_\text{gas} = 5.82 \times 10^5\,\Msun$, respectively. The gravitational forces are softened on scales of \mbox{$2.2$\,ckpc} with the smallest cell sizes reaching $10$\,pc. This allows us to model atomic cooling haloes throughout the entire simulation volume. (See \citet{Smith2022} for additional discussion showing that \thesan has sufficient resolution to accurately model Ly$\alpha$ intrinsic emission and IGM transmission, but that there would be significant uncertainties in the emergent spectra resulting from Monte Carlo Ly$\alpha$ radiative transfer calculations.)

\thesan also includes a suite of medium resolution simulations with the same initial conditions aiming to investigate the impacts on reionization caused by different physics as described in \citet{Kannan2022}. The \thesantwo simulation is the same as \thesanone but with two (eight) times lower spatial (mass) resolution. The \thesanwc weak convergence simulation slightly increases the birth cloud escape fraction to compensate for lower star formation in the medium resolution runs. The \thesansdao assumes an alternative dark matter model that includes couplings to relativistic particles giving rise to strong Dark Acoustic Oscillations (sDAOs) cutting off the linear matter power spectrum at small scales. The \thesanhigh (\thesanlow) simulations use a halo-mass-dependent escape fraction, with only haloes above (below) $10^{10}\,\Msun$ contributing to reionization. We note that to match the observed neutral hydrogen fraction, the different \thesan runs adopt different birth cloud escape fractions $f_\text{esc}^\text{cloud}$: \thesanone and \thesantwo have $0.37$, \thesanwc has $0.43$, \thesanhigh has $0.8$, \thesanlow has $0.95$, and \thesansdao has $0.55$. The reionization histories of each simulation are plotted in Fig.~\ref{fig:reion_history}, where \thesanlow reionizes faster due to the greater contribution of low-mass galaxies, with the opposite being true of \thesanhigh. \thesanwc is also closer than \thesantwo to the reionization history of \thesanone. The \thesan simulations have been used to study 21\,cm power spectra \citep{Kannan2022} including an effective bias expansion in redshift space \citep{Qin2022}, IGM--galaxy connections including ionizing mean free path and Ly$\alpha$ transmission statistics \citep{Garaldi2022}, multitracer line intensity mapping \citep{KannanLIM2022}, and galaxy ionizing escape fractions \citep{Yeh2023}.

\begin{figure*}
  \centering
  \includegraphics[width=\textwidth]{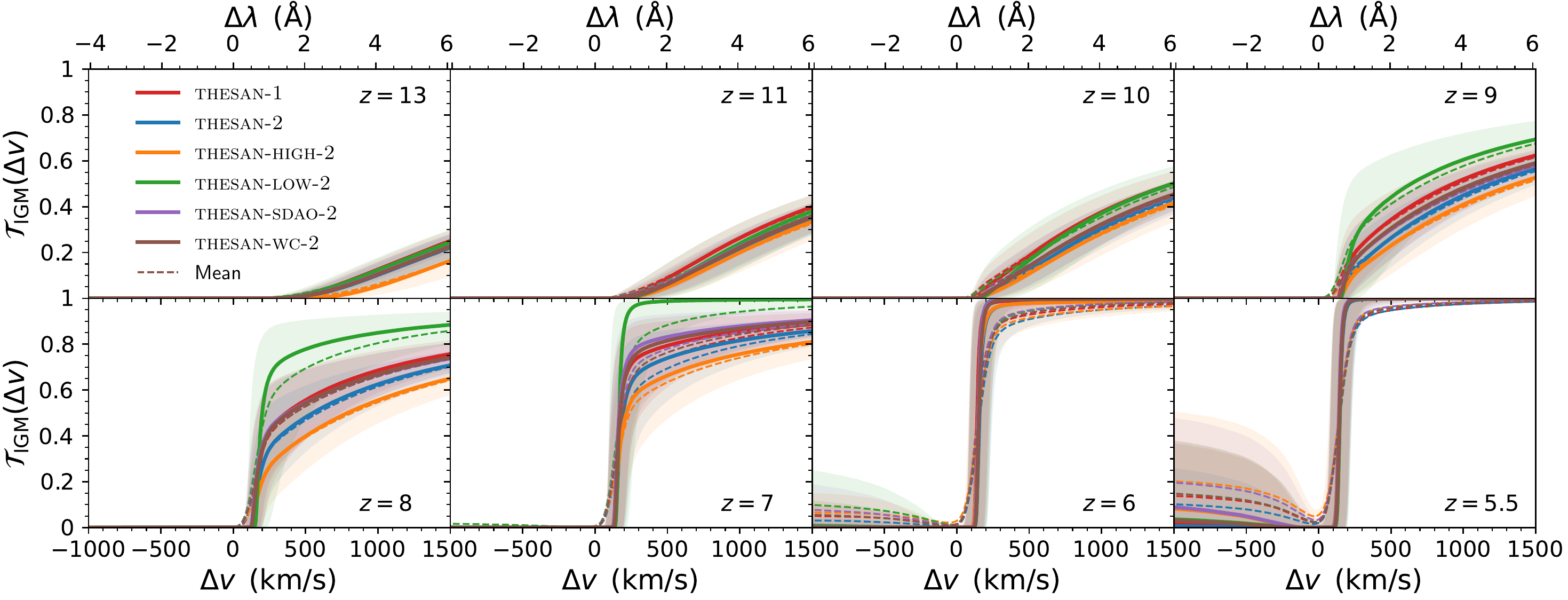}
  \caption{IGM transmission $\mathcal{T}_\text{IGM}$ as a function of velocity offset $\Delta v$ and rest-frame wavelength offset $\Delta \lambda$ around the Ly$\alpha$ line for each simulation. Each panel shows a different redshift over $z \in [5.5,13]$. The solid (dashed) curves show the catalogue median (mean) statistics and shaded regions give the $1\sigma$ confidence levels. With equal weight to all galaxies and sightlines, this view is biased towards more numerous lower-mass haloes. From redshifts $z \approx 7$--$9$, the various simulations produce higher or lower transmission curves as a result of the different reionization histories and bubble morphologies. The shapes of coeval curves are also affected, e.g. with steeper or flatter slopes around $\Delta v \approx 500\,\text{km\,s}^{-1}$. In order of decreasing transmissivity these are \thesanlow, \thesansdao, \thesanwc, \thesanone, \thesantwo, and \thesanhigh.}
  \label{fig:T_IGM_catalog}
\end{figure*}

\subsection{Ly\texorpdfstring{$\balpha$}{α} emission and transmission}
\label{sec:Lya_methods}
We summarize the methodology described in \citet{Smith2022} for Ly$\alpha$ emission and IGM transmission calculations, noting that we now include all \thesan simulations. For each snapshot, we produce a Ly$\alpha$ catalogue directly mirroring the friends-of-friends (FoF) halo and \textsc{subfind} subhalo catalogues. This provides supplemental data for dark matter haloes and gravitationally bound galaxies, as well as high level information such as the global emissivity. For each galaxy we track the total Ly$\alpha$ luminosity $L_\alpha$\,[$\text{erg\,s}^{-1}$] including contributions from resolved recombination $L_\alpha^\text{rec}$, collisional excitation $L_\alpha^\text{col}$, and unresolved \HII regions $L_\alpha^\text{stars}$, stellar continuum spectral luminosities $L_{\lambda,\text{cont}}$\,[$\text{erg\,s}^{-1}\text{\AA}^{-1}$] at $\lambda_\text{cont} = \{1216, 1500, 2500\}\,\text{\AA}$, ionizing luminosity from active galactic nuclei $L_\text{ion}^\text{AGN}$\,[$\text{erg\,s}^{-1}$], and centre of Ly$\alpha$ luminosity position $\bmath{r}_\alpha$\,[kpc], peculiar velocity $\bmath{v}_\alpha$\,[$\text{km\,s}^{-1}$], and 1D velocity dispersion $\sigma_\alpha$\,[$\text{km\,s}^{-1}$]. The details for calculating the total intrinsic emission due to recombinations, collisional excitation, and local stars are described in \citet{Smith2022} (see their equations~1--3), which we summarize in a single equation as:
\begin{equation} \label{eq:L_stars}
  \frac{L_\alpha}{h \nu_\alpha} = \int \left[ P_\text{B} \alpha_\text{B} n_p + q_\text{col} n_\text{\HI} \right] n_e \text{d}V + 0.68 (1 - f_\text{esc}^\text{cloud}) \dot{N}_\text{ion} \, .
\end{equation}
Here $h \nu_\alpha = 10.2\,\text{eV}$, $P_\text{B}$ is the Ly$\alpha$ conversion probability per recombination event, $\alpha_\text{B}$ is the case B recombination coefficient, $q_\text{col}$ is the collisional rate coefficient, and $n_p$, $n_\text{\HI}$, and $n_e$ are number densities for protons, neutral hydrogen, and electrons. Finally, $0.68$ is the fiducial conversion probability, $f_\text{esc}^\text{cloud}$ denotes the escape fraction of ionizing photons calibrated for each simulation to match reionization history constraints, and $\dot{N}_\text{ion}$ is the age and metallicity dependent emission rate of ionizing photons from stars taken from the BPASS models (v2.2.1). Combining resolved and sub-resolution emission mechanisms is necessary given that $f_\text{esc}^\text{cloud} < 1$. Moreover, this approach allows us to utilize the on-the-fly ionization field while also reproducing the expected ionizing photon budget.

We also produce catalogues for the frequency-dependent transmission of Ly$\alpha$ photons through the IGM for all simulations at $z = \{5.5, 6, 6.6, 7, 8, 9, 10, 11, 13\}$ \citep[similar to][]{Laursen2011,Byrohl2020,Gronke2021,Garel2021,Park2021}. As in \citet{Smith2022}, for each central halo selected with at least 32 star particles (and less resolved ones if they are among the 90\% most massive centrals) we extract 768 radially outward rays corresponding to equal area healpix directions of the unit sphere. We use the COsmic Ly$\alpha$ Transfer code \citep[\textsc{colt};][]{Smith2015,Smith2019,Smith2022MW} to perform exact ray-tracing through the native Voronoi unstructured mesh data. We start the rays at initial distances of $R_\text{vir}$, defined as the radius within which the mean density becomes 200 times the cosmic value ($R_{200}$). We take the systemic location and rest-frame from the subhalo intrinsic Ly$\alpha$ luminosity averaged position $\bmath{r}_\alpha$ and velocity $\bmath{v}_\alpha$. We select a broad wavelength range of $\Delta v \in [-2000, 2000]\,\text{km\,s}^{-1}$ sampled at a high spectral resolution of $5\,\text{km\,s}^{-1}$ or a resolving power of $R \approx 60\,000$. We perform the integrations out to a distance of $4000\,\text{km\,s}^{-1} / H(z) \approx 40\,\text{cMpc}\,[(1+z)/7]^{-1/2}$. We calculate the traversed optical depth based on the continuous Doppler shifting scheme described in \citet{Smith2022}, which incorporates velocity gradients encountered during propagation. The total optical depth $\tau$ defines the frequency-dependent transmission function for each ray:
\begin{equation}
  \mathcal{T}_\text{IGM}(\Delta v) \equiv \exp\big[ -\tau(\Delta v) \big] \, ,
\end{equation}
which describes the fraction of flux not attenuated by the IGM after escaping the halo. Similar to \citet{Park2021}, we incorporate the distant IGM beyond the local rays in a statistical sense based on the global reionization history. This additional damping-wing absorption $\tau_\text{DW}$ has a very minor impact after the EoR but is increasingly important at higher redshifts \citep{Miralda-Escude1998}. Finally, we adopt the complete first-order quantum-mechanical correction to the Voigt profile presented by \citet{Lee2013}, which strengthens the red wing due to positive interference of scattering from all other levels. Detailed calculations for $\tau_\text{DW}$ and the Voigt profile correction are given in \citet{Smith2022}.

While \thesan is fully capable of capturing IGM scale effects, detailed LAE modelling is strongly affected by ISM scale sourcing and radiation transport through the CGM. The uncertainties mainly arise from the galaxy formation model (and limited resolution), which includes temporary decoupling of wind particles from the hydrodynamics and the use of an effective equation of state (EoS) for cold gas above the density threshold $n_\text{H} \approx 0.13\,\text{cm}^{-3}$ \citep{SpringelHernquist2003}. In this study, we choose to calibrate an empirical model for the spectral line profile emerging from galaxies and absorption of Ly$\alpha$ photons by dust (see Sec.~\ref{sec:calibration}). We emphasize that due to dust and IGM reprocessing the bright end of the observed Ly$\alpha$ LF differs from the intrinsic one by up to two orders of magnitude. Thus, our approach informs us about what is required to match observations and provides observed LAE luminosity catalogues from the \thesan simulations for follow-up comparison studies. In Fig.~\ref{fig:visualization}, we show a preview visualization of the LAE catalogue along with the resolved recombination rate density (for details see Section~\ref{sec:implications}).

\section{Transmission statistics}
\label{sec:transmission}
\begin{figure}
  \centering
  \includegraphics[width=\columnwidth]{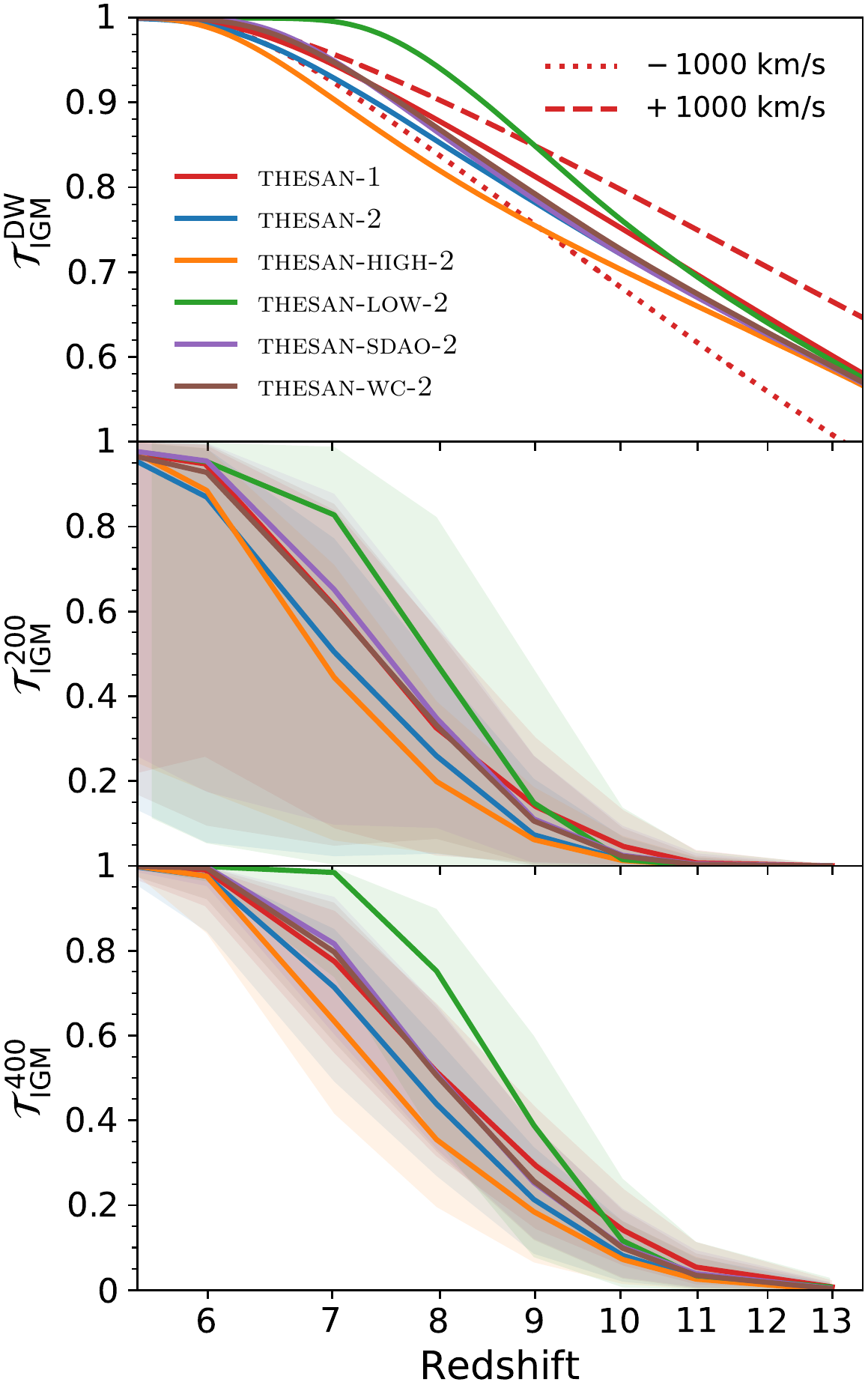}
  \caption{\textit{Top:} Damping-wing IGM transmission $\mathcal{T}_\text{IGM}^\text{DW} = \exp(-\tau_\text{DW})$ at $\Delta v = 0$ for all simulations as a function of redshift. For~\thesanone we also plot $\mathcal{T}_\text{IGM}^\text{DW}$ at $\Delta v = \pm 1000\,\text{km\,s}^{-1}$ to indicate the range of frequency dependence. \textit{Middle and bottom:} Overall transmission averaged over $50\,\text{km\,s}^{-1}$ spectral windows at velocity offsets of $\Delta v = 200$ and $400\,\text{km\,s}^{-1}$ plotted over redshift, which are qualitatively similar, but $\mathcal{T}_\text{IGM}^{400}$ has slightly higher transmission and less variation. The effects of different reionization histories can be seen again with \thesanlow and \thesanhigh bracketing the other simulations.}
  \label{fig:DW_200_400}
\end{figure}

\begin{figure}
  \centering
  \includegraphics[width=\columnwidth]{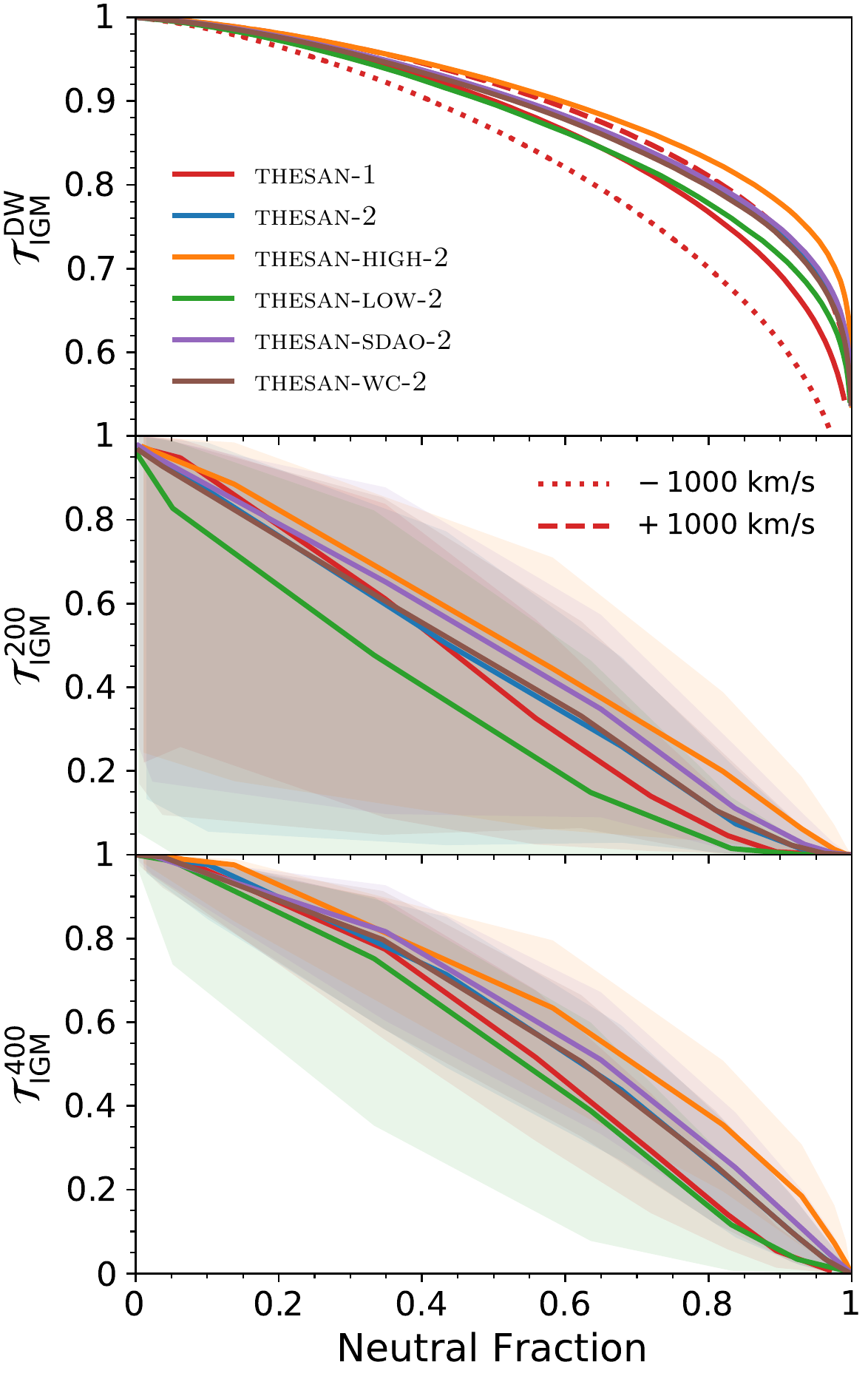}
  \caption{The same quantities as in Fig.~\ref{fig:DW_200_400}, i.e. the IGM transmission from the damping wing $\mathcal{T}_\text{IGM}^\text{DW}$ and at velocity offsets of $\Delta v = 200$ and $400\,\text{km\,s}^{-1}$ (from top to bottom), but as functions of the global neutral fraction. In this view the transmission order for \thesanlow and \thesanlow have switched places, illustrating the importance of bubble morphology in addition to the reionization history.}
  \label{fig:DW_xHI_200_400}
\end{figure}

\begin{figure*}
  \centering
  \includegraphics[width=\textwidth]{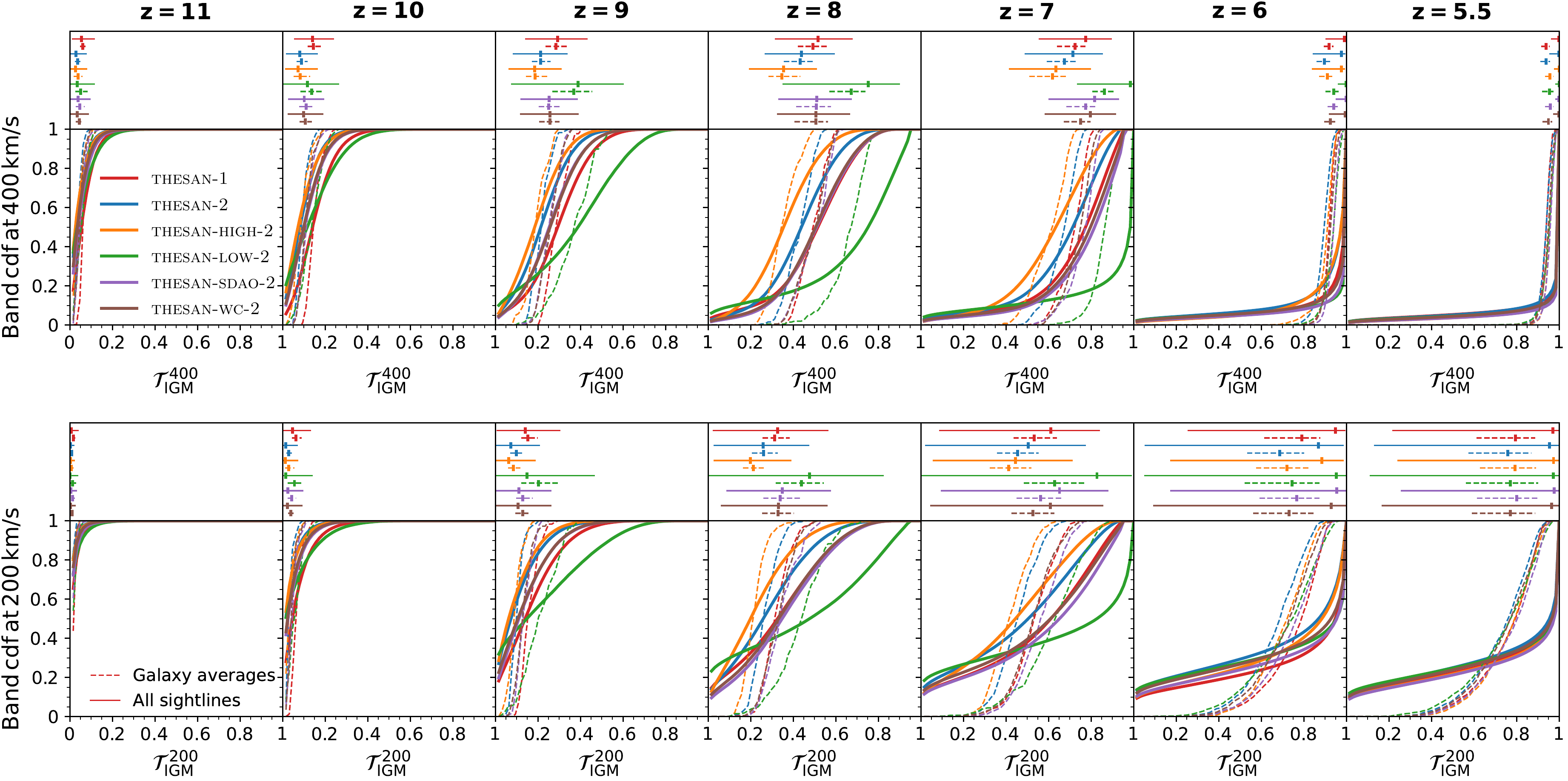}
  \caption{Cumulative probability distribution functions (CDF) for transmission at $200$ and $400\,\text{km\,s}^{-1}$ for each simulation at several redshifts. Median and 1$\sigma$ summary statistics are plotted in the upper panels. We also include each CDF of galaxy averages plotted as dashed curves, which result in slightly lower medians and have narrower distributions. Thus, the variation across sightlines from a single LAE is larger than the variation between different galaxies.}
  \label{fig:T_IGM_cdfetc}
\end{figure*}

The strong absorption of photons near the Ly$\alpha$ line provides a powerful probe of the structure and evolution of reionization. In this section we explore the isolated impact of IGM transmission in \thesan. We note that we apply a UV continuum selection cut of $M_{1500} < -19$ when aggregating haloes for summary statistics to avoid resolution biases in comparisons of different simulations (see Figs.~\ref{fig:T_IGM_catalog}--\ref{fig:T_IGM_cdfetc}). Due to the low number of bright galaxies we relax this restriction to $M_{1500} < \{-18.5,-17.5,-16.5,-15.5\}$ for $z = \{9,10,11,13\}$, respectively.

\subsection{Comparing transmission curves}
We first consider the frequency dependence and redshift evolution of global catalogue statistics. In Fig.~\ref{fig:T_IGM_catalog} we show the IGM transmission fraction $\mathcal{T}_\text{IGM}$, including the local and damping-wing contributions as a function of velocity offset $\Delta v$ and rest-frame wavelength offset $\Delta \lambda$ around the Ly$\alpha$ line for each simulation. Each panel shows the transmission at a different redshift across the range $z \in [5.5,13]$. The solid (dashed) curves show the full catalogue median (mean) statistics and shaded regions give the $1\sigma$ confidence levels. With equal weight to all galaxies and sightlines, this view is biased towards lower masses that dominate the catalogues by halo count. As expected, transmission redward of line centre increases as redshift decreases due to the decreasing neutral fraction, while the blue peak is highly suppressed until the end of the simulation. In Appendix~\ref{appendix:resolution}, we explore the impact of IGM resolution on the frequency-dependent transmission curves and conclude that while there are strong differences near line centre, the difference in the red peak transmission ($\Delta v \gtrsim 200\,\text{km\,s}^{-1}$) is relatively minor.

From redshifts $z \approx 7$--$9$, the various simulations can produce significantly higher or lower transmission curves as a result of the different reionization histories and bubble morphologies with \thesanlow and \thesanhigh bracketing the range. The shapes of coeval transmission curves are also affected; for example, \thesanlow looks more like a step function, and around a velocity offset of $\Delta v \approx 500\,\text{km\,s}^{-1}$ the other simulations have steeper slopes. \thesanlow has the earliest global reionization history but also the greatest advantage for local transmission around the abundant low-mass galaxies, as these dominate reionization in this scenario. In contrast, \thesanhigh has the lowest degree of transmission at all redshifts, as a consequence of having the latest reionization history and mainly providing a local transmissivity boost to rarer, higher-mass galaxies.

\subsection{Redshift evolution}
To highlight the rapid change in transmissivity throughout the EoR, we focus on the behaviour in wavelength bands relevant for Ly$\alpha$ related science. Specifically, we define the integrated or mean transmission as $\mathcal{T}_\text{IGM}^\text{\,int} \equiv \int \mathcal{T}_\text{IGM}\,\text{d}\Delta v / \int \text{d}\Delta v$, focusing on velocity offset windows centred at $200$ and $400\,\text{km\,s}^{-1}$ averaged over widths of $50\,\text{km\,s}^{-1}$ to match typical spectroscopic instrument capabilities. These locations are chosen to roughly correspond to the left and right sides of observed red peaks at low and high redshifts \citep[e.g.][]{Ouchi2020}. In Fig.~\ref{fig:DW_200_400} we plot the damping wing transmission at line center as a function of redshift in the top panel, and the transmission at $\Delta v = 200$ and $400\,\text{km\,s}^{-1}$ with redshift in the bottom two panels. In the top panel, we also plot damping wing transmission at $\Delta v = \pm 1000\,\text{km\,s}^{-1}$ for \thesanone to demonstrate the range of dependence on frequency; there is higher transmission for red velocity offsets, and vice versa. At $\Delta v = 400\,\text{km\,s}^{-1}$, there is less variation within and among the simulations than at $200\,\text{km\,s}^{-1}$. We interpret this as a difference between probing the distant absorption across a more homogeneous large-scale IGM and the nearby imprint of the stochastic local environment. The damping wing transmission in \thesanone is generally higher than the other simulations, with the exception of \thesanlow, due to its higher resolution that allows more low-mass galaxies to contribute ionizing photons and influence the early reionization history. However, the overall transmission in \thesanone typically lies between the \thesantwo and \thesanwc simulations. We note that the $8$ times higher resolution of \thesanone is expected to induce systematically lower transmission because the covering fraction of Lyman-limit systems increases \citep{vandeVoort2019}.

The effects of the different reionization histories can be seen again, with \thesanlow bracketing the simulations from above and \thesanhigh from below in all panels. To disentangle the effects of differing reionization histories, in Fig.~\ref{fig:DW_xHI_200_400} we also plot the same values against the global volume-weighted neutral fraction. In this perspective the transmission fractions essentially increase linearly with decreasing neutral fraction, while damping wing transmission increases faster at higher neutral fractions before leveling off. It is also interesting that \thesanlow (\thesanhigh) now has lower (higher) transmission than the other simulations, switching places and highlighting the importance of bubble morphology in addition to the reionization history; \thesanlow (\thesanhigh) likely allows less (more) transmission at the same ionization fraction due to its smaller (larger) bubble sizes. In addition, \thesansdao and \thesanwc clearly have higher transmission than \thesanone at the same neutral fraction, especially at larger neutral fractions. This is also due to smaller bubble sizes in \thesanone at higher redshifts during first formation, enabled by its higher resolution \citep[see section 3.4 of][and Neyer et al. in prep.]{Kannan2022}. Otherwise, \thesansdao and \thesanwc trend closely to \thesanone in other statistics. We note that the damping wing contribution has less variation between simulations due to the assumption of a homogeneous cosmological neutral hydrogen density, removing information about morphology but retaining an imprint of the timing of reionization. As before, both Fig.~\ref{fig:DW_200_400} and Fig.~\ref{fig:DW_xHI_200_400} are biased towards lower mass haloes.

\begin{figure*}
  \centering
  \includegraphics[width=\textwidth]{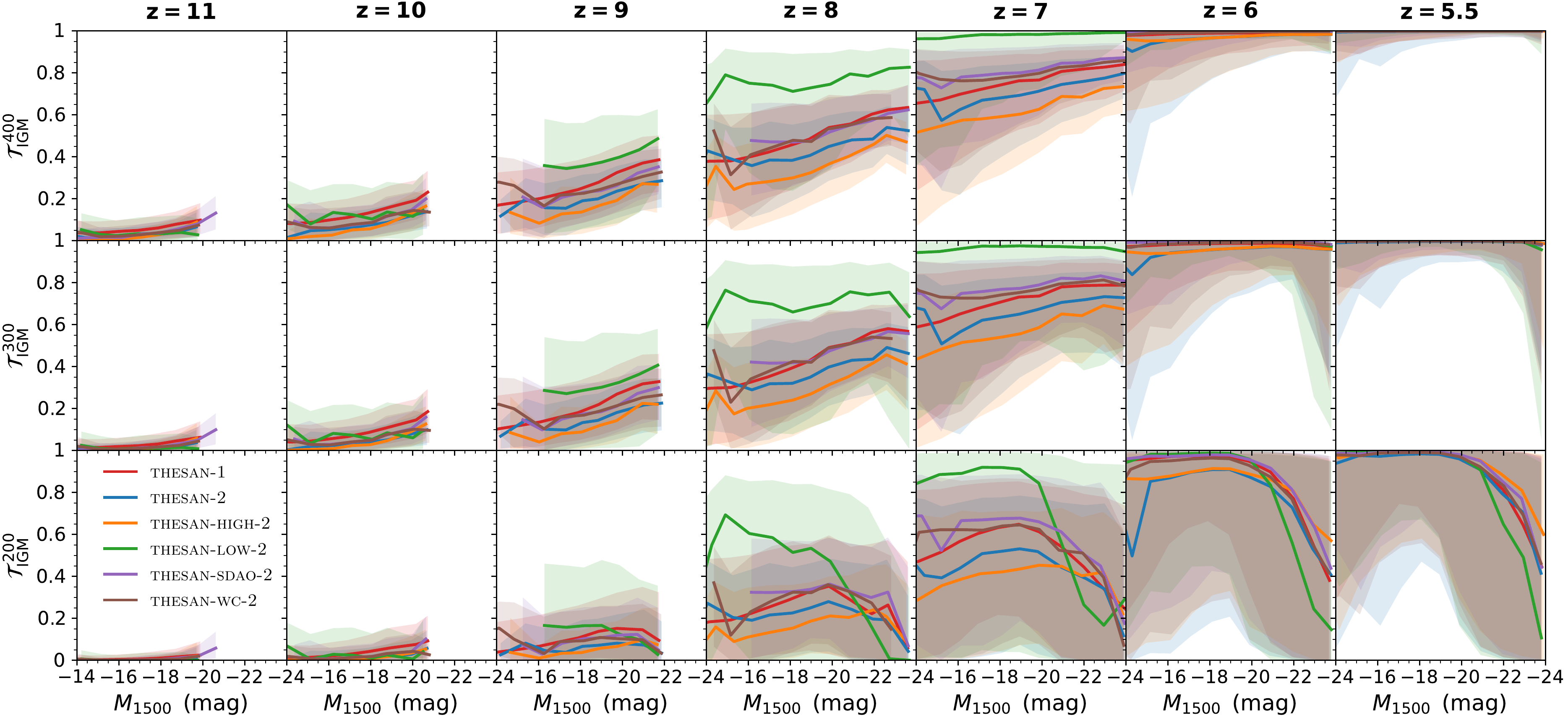}
  \caption{Band integrated IGM transmission median and 1$\sigma$ statistics as a function of UV magnitude $M_{1500}$ for each redshift. We find clear trends of higher transmission for brighter galaxies, especially for $\Delta v \gtrsim 300\,\text{km\,s}^{-1}$ at $z \gtrsim 6$. Another striking feature is the strong absorption for bright galaxies ($M_{1500} \lesssim -20$) due to infalling gas at $\Delta v \lesssim 200\,\text{km\,s}^{-1}$.}
  \label{fig:T_IGM_M1500}
\end{figure*}

\subsection{Transmission distributions}
To further investigate the different distributions of transmission between simulations, in Fig.~\ref{fig:T_IGM_cdfetc} we plot the cumulative probability distributions for transmission at $\Delta v = 200$ and $400\,\text{km\,s}^{-1}$ for each simulation at several redshifts. For convenience in comparing simulations, the median values and 1$\sigma$ summary statistics are plotted in the upper panels. This view is again biased towards lower masses. The solid curves indicate the distribution for all sightlines across all galaxies, while for the dashed lines, we first average the transmission for each galaxy over its 768 sightlines before plotting the distribution for all galaxies. Generally, the galaxy-averaged transmission is slightly lower and has a more narrow distribution, especially at lower redshifts. This implies that the variation across sightlines from a single LAE is larger than the variation between different galaxies. This arises from the exponential sensitivity on optical depth ($e^{-\tau}$) allowing the same galaxy to have both large and small values; i.e. the broad and bimodal sightline distributions are not due to different large-scale IGM environments. There will always be sightlines with very low transmission due to filaments and other self-shielding structures common at high-$z$ (\citealp{Park2021}; healpix plots from \citealp{Smith2022}).

The distributions at $\Delta v = 200\,\text{km\,s}^{-1}$ are much broader than those at $400\,\text{km\,s}^{-1}$, again indicating that it has more variation in transmission. This is both in terms of values among galaxies in the same simulation and between different simulations (recall the shaded regions in Fig.~\ref{fig:DW_200_400}). There are also lower median transmission values at $200\,\text{km\,s}^{-1}$ at a given redshift because the closer photons are to line center, the more susceptible they are to resonant infalling motions within the local environments of galaxies, which vary significantly.

In addition, \thesanone, with its higher resolution, is able to resolve lower-mass galaxies, and \thesanlow has lower-mass galaxies dominate reionization; the two simulations therefore have earlier ionized bubble formation around lower mass galaxies. This leads to the higher transmission values that we initially see at $z \gtrsim 10$. Eventually, \thesanlow diverges as it has a much earlier completion of reionization, while \thesanhigh lags behind the others due to its later reionization history. Overall, the similar behaviour of these statistics despite changing resolution and physics demonstrates that Ly$\alpha$ transmission does indeed act as a powerful probe of the EoR.

\begin{figure*}
  \centering
  \includegraphics[width=\textwidth]{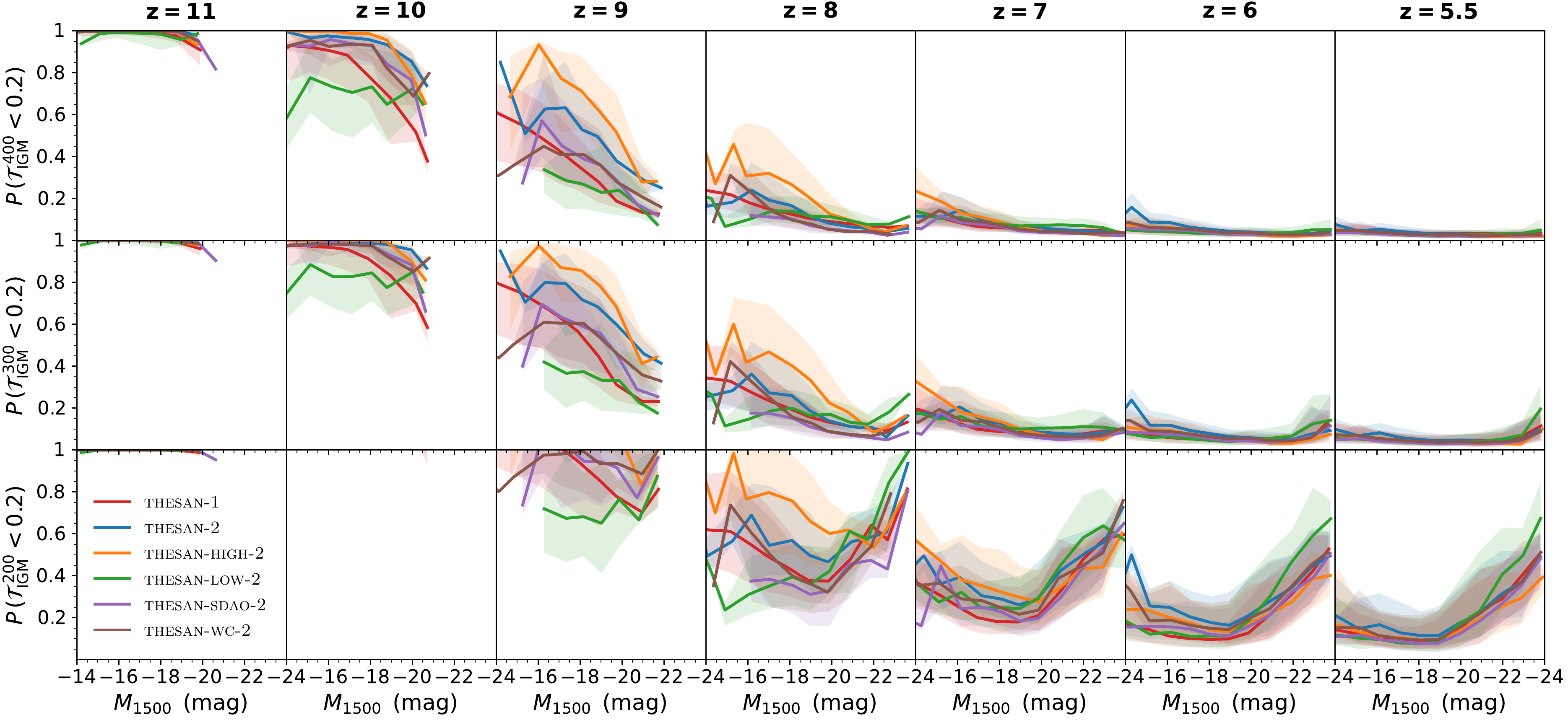}
  \caption{IGM transmission covering fractions defined as the fraction of sightlines around each galaxy with transmission below 20 per cent, i.e. $P(\mathcal{T}_\text{IGM} < 0.2)$. The curves and shaded regions give the median and 1$\sigma$ variation as a function of UV magnitude $M_{1500}$ at velocity offsets of $\Delta v = \{200, 300, 400\}\,\text{km\,s}^{-1}$ (bottom to top) at a range of redshifts (decreasing from left to right). See the text for additional discussion.}
  \label{fig:covering_fractions}
\end{figure*}

\subsection{Dependence on UV magnitude}
We now explore the dependence of transmission on galaxy properties to quantify environmental effects and compare similar galaxies across different simulations. In Fig.~\ref{fig:T_IGM_M1500}, we plot the median transmission for $50\,\text{km\,s}^{-1}$ wavebands centred at velocity offsets of $\Delta v = \{200, 300, 400\}\,\text{km\,s}^{-1}$ as functions of UV magnitude $M_{1500}$ (without any dust correction) at several redshifts with shaded regions indicating 1$\sigma$ statistics. Transmission generally increases with magnitude, with the notable exception of a downturn for the brightest galaxies ($M_{1500} \lesssim -20$) at $\Delta v = 200\,\text{km\,s}^{-1}$. This effect also bleeds into the transmission at $300\,\text{km\,s}^{-1}$, but disappears by $400\,\text{km\,s}^{-1}$. This suppression signature is mainly due to cosmic streams of infalling gas in the crowded environments around massive galaxies, as the frequency range likely to encounter a resonance point is extended to the circular velocity $V_c \propto M_\text{halo}^{1/3}$ \citep{Santos2004,Dijkstra2007}.

We again see \thesanlow and \thesanhigh bracketing the other simulations. It is also interesting that although \thesanlow generally has higher transmission, its typical transmission at $200\,\text{km\,s}^{-1}$ dips below the other simulations in the downturn in the brightest galaxies, while \thesanhigh goes above the others. This is due to the biased (dis)advantage artificially setting the birth cloud ionizing escape fractions to be (zero) higher in the (\thesanlow) \thesanhigh runs. The remaining simulation orderings are consistent with differences in the reionization histories. In practice, due to the large variances it may be difficult to disentangle timing, morphology, and modelling effects from IGM transmission alone.

\subsection{Covering fractions}
Covering fractions are often used to understand stochastic visibility arising from anisotropic and bimodal sightline distributions. In our context we define the covering fraction of each individual galaxy as the fraction of sightlines with transmission below 20 per cent; i.e. $P(\mathcal{T}_\text{IGM} < 0.2)$. In Fig.~\ref{fig:covering_fractions} we show the median and 1$\sigma$ range of halo covering fractions as a function of UV magnitude $M_{1500}$ for each simulation at different redshifts and wavebands. Overall, the covering fractions decrease with decreasing redshift, as the progression of reionization determines the amount of neutral hydrogen around these haloes. We also see that fainter galaxies have larger covering fractions, corresponding to more isotropic suppression due to the correlation with ionized bubble size. We also see the influence of infall velocity absorption and local environment with an upturn for bright galaxies ($M_{1500} \lesssim -20$) at $\Delta v = 200\,\text{km\,s}^{-1}$, which is largely absent by $400\,\text{km\,s}^{-1}$. Finally, lower (higher) transmission fractions translate to higher (lower) covering fractions, so the reverse ordering among simulations is also largely preserved.

\section{Observed luminosity functions}
\label{sec:LF}
Observed Ly$\alpha$ LFs are expected to differ significantly from intrinsic ones as a result of dust absorption and flux-averaged IGM transmission. In \thesan we can accurately model the Ly$\alpha$ intrinsic emission from galaxies and frequency-dependent IGM transmission, but ISM scale Ly$\alpha$ radiative transfer effects would be extremely challenging to model self-consistently. The empirical calibration approach presented in this section is intended as a reasonable alternative to sub-grid model dependent predictions, and provides a bridge framework to incorporate emission, dust, spectra, and transmission modelling components to match observed Ly$\alpha$ LFs.

\begin{table}
  \centering
  \caption{Summary of various emergent spectra models used in previous studies. The columns from left to right are the acronym used in the text and figures, reference citation, model description, and relevant notes about the simulation or scientific application. These models are provided for comparison with our fully calibrated spectral model shown in Fig.~\ref{fig:nodustcomparison}, both as purely spectral models and also including our best-fit dust model. For notational compactness $G(\mu,\sigma)$ denotes a Gaussian with mean $\mu$ and standard deviation $\sigma$.}
  \label{tab:catalog}
  \begin{tabular}{ccl}
  \hline
  Model & Reference & Description \\
  \hline
  J13 & \citet{Jensen2013} & Gaussian-minus-a-Gaussian$^a$ \\
  W19 & \citet{Weinberger2019} & $G(\{1.5, 1.8\}\,V_c, 88\,\text{km\,s}^{-1})^b$ \\
  G21 & \citet{Gangolli2021} & $G(\{1.2, 1.4, 1.8\}\,V_c, V_c)^c$ \\
  BG20G & \citet{Byrohl2020} & $G(0, 200\,\text{km\,s}^{-1})$ \\
  BG20N & \citet{Byrohl2020} & Neufeld double-peaked profile$^d$ \\
  BG20R & \citet{Byrohl2020} & Neufeld with red peak only \\
  \hline
  \multicolumn{3}{l}{$^a$ The $\sigma$ values depend on halo mass, calibrated to simulated lines} \\
  \multicolumn{3}{l}{$^b$ The $\mu$ coefficient depends on their reionization model} \\
  \multicolumn{3}{l}{$^c$ The $\mu$ coefficient depends on neutral fraction and redshift} \\
  \multicolumn{3}{l}{$^d$ Temperature $T = 10^{4}\,\text{K}$ and \HI column density $N_{\rm HI} = 10^{20}\,\text{cm}^{-2}$}
  \end{tabular}
\end{table}

\begin{figure}
  \centering
  \includegraphics[width=\columnwidth]{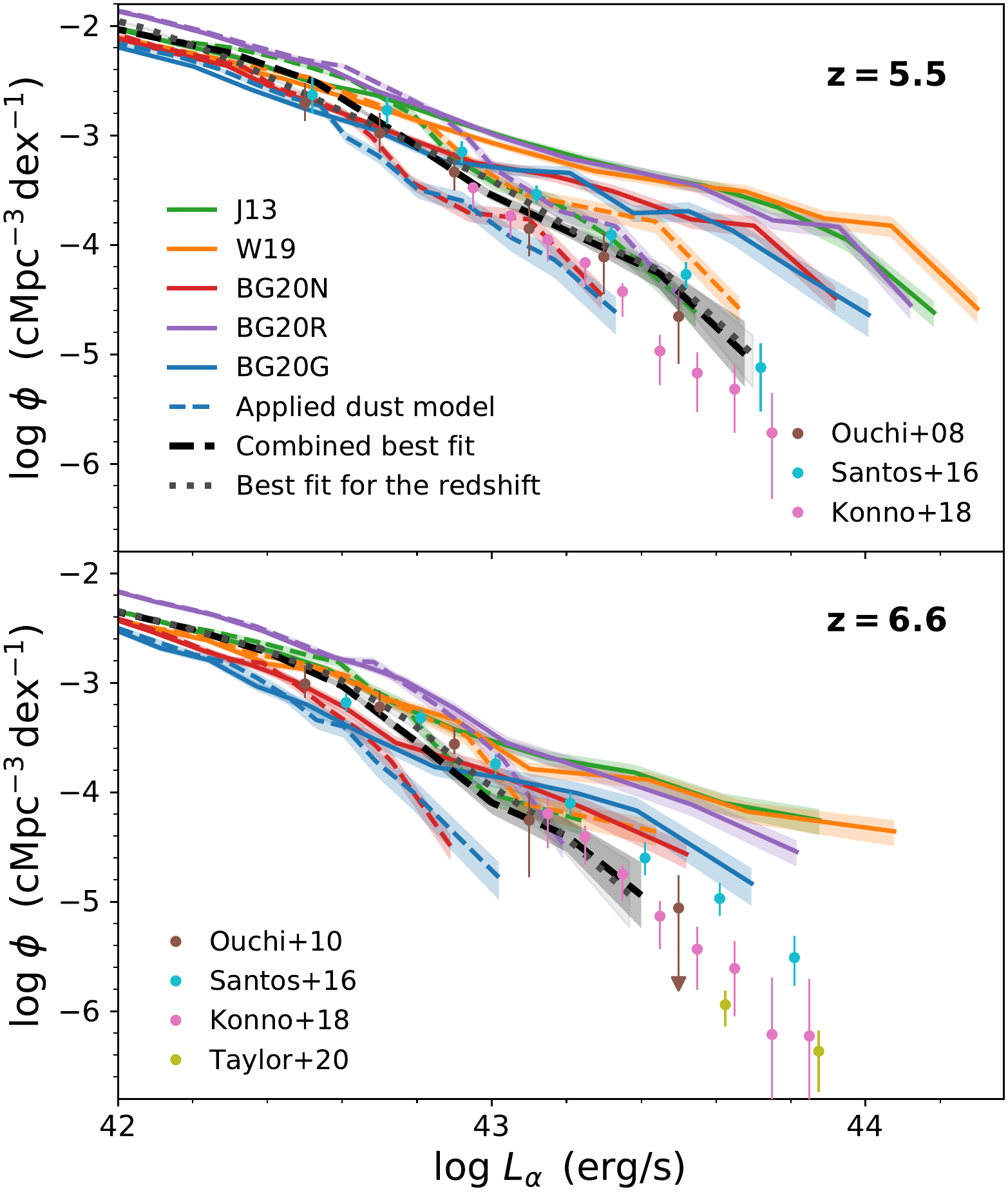}
  \caption{Comparison of observed Ly$\alpha$ luminosity functions at $z=5.5$ (top panel) and $z=6.6$ (bottom panel) for the spectral models described in Table~\ref{tab:catalog}, which do not include dust modelling, as applied to the \thesanone simulation (coloured solid curves; see the table for a mapping between identifier and model). For comparison, we show a collection of observational data in the background. The thick black dashed curve shows the best-fitting model with the same parameters used for both redshifts, with the shaded regions showing the variation in the LF across sightlines. The grey dotted curve shows the case where we only fit to the data for the specific redshift, with our calibrated models showing significantly better agreement with the observed data than the uncalibrated spectral models already available in the literature, even after also applying our dust model (coloured dashed curves). For the W19 model we chose 1.5 as the coefficient for $\mu$.}
  \label{fig:nodustcomparison}
\end{figure}

\subsection{Galaxy model calibration procedure}
\label{sec:calibration}
The choice of spectral model can significantly alter the predicted observed Ly$\alpha$ luminosities of galaxies. In Table~\ref{tab:catalog}, we provide a summary of previous simulation-based studies from across the literature that adopt empirical emergent Ly$\alpha$ spectra models. We also assign a name abbreviation to each reference and give a brief summary of relevant information. In Fig.~\ref{fig:nodustcomparison} we apply these models to our \thesanone IGM transmission data at redshifts of $5.5$ and $6.6$ (we note that we do not show the model from \citet{Gangolli2021}, as it overlaps with the curve from \citet{Weinberger2019}), alongside two additional curves representing our best-fit spectral and dust attenuation model at the specific redshift (dotted black) and when combining redshifts (thick dashed black) for which the calibration process is discussed in detail below. We see that none of the currently available models in the literature (coloured curves), when applied to the \thesanone simulation, provide an acceptable fit to the observed data (background points with error bars). This is especially problematic at the highest luminosities because none of the models accounts for dust absorption. This highlights the importance of including dust in addition to IGM transmission, which strongly suppresses more massive galaxies that are intrinsically bright but have low Ly$\alpha$ escape fractions. We emphasize that when we apply our dust model (detailed below) to the spectral models from the literature, the agreement with observations improves but not to a satisfactory level unless both models are simultaneously constrained.

\begin{figure}
  \centering
  \includegraphics[width=\columnwidth]{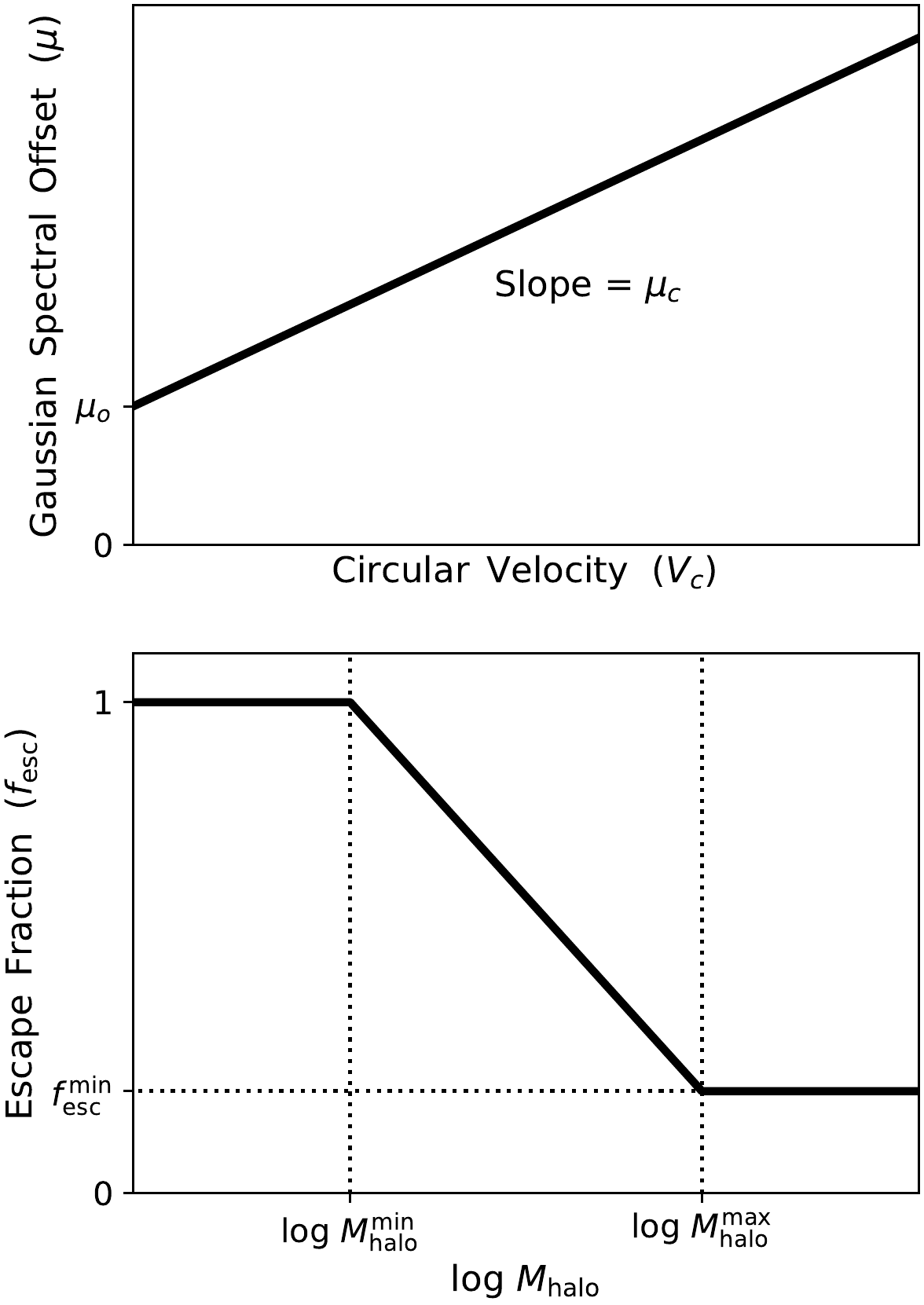}
  \caption{Schematic representations for the adopted model of the emergent Gaussian spectral offset $\mu$ as a function of halo circular velocity $V_c$ (upper panel; see equation~\ref{eq:mu}) and Ly$\alpha$ escape fraction $f_\text{esc}$ due to dust absorption as a function of halo mass $M_\text{halo}$ (lower panel; see equation~\ref{eq:f_esc}).}
  \label{fig:modelplots}
\end{figure}

Unfortunately, there are aspects of parameterised dust models that are degenerate with accompanying spectral models; for example, strong dust absorption will lower the observed Ly$\alpha$ luminosity, but so will a spectral model with very little flux in the red damping wing. To account for this, we choose to model both the dust and galaxy spectra for completeness. Fig.~\ref{fig:modelplots} contains a schematic representation of both our dust and spectral models, which we now describe.

For simplicity, we choose to model dust as a piecewise function of galaxy mass. Below a minimum halo mass, $M_\text{halo}^\text{min}$, the Ly$\alpha$ escape fraction due to dust stays constant at unity, i.e. we assume negligible absorption. Above a maximum halo mass, $M_\text{halo}^\text{max}$, the escape fraction stays constant at a certain minimum escape fraction, $f_\text{esc}^\text{min}$. In the intermediate mass range, the escape fraction decreases log-linearly. The free parameters are thus the maximum and minimum halo masses, and the minimum escape fraction:
\begin{equation} \label{eq:f_esc}
  f_\text{esc} =
    \begin{cases}
      1 & M_\text{halo} < M_\text{halo}^\text{min} \\
      f_\text{esc}^\text{min} & M_\text{halo} > M_\text{halo}^\text{max} \\
      1 + (f_\text{esc}^\text{min} - 1) \frac{(\log M_\text{halo} - \log M_\text{halo}^\text{min})}{(\log M_\text{halo}^\text{max} - \log M_\text{halo}^\text{min})} & \quad \text{otherwise}
    \end{cases} \, .
\end{equation}
Likewise, our spectral model is chosen to be a Gaussian line profile described by a constant width of $\sigma = 200\,\text{km\,s}^{-1}$. We chose a constant line width as we found that varying the line width was strongly degenerate with changes in the modelled minimum escape fraction. Our spectral model is given by
\begin{equation} \label{eq:J}
  J(\nu) \propto \text{exp}\left(-\frac{1}{2}\,\frac{(\Delta v - \mu)^2}{\sigma^2}\right) \, ,
\end{equation}
with
\begin{equation} \label{eq:mu}
    \mu = \mu_c\,V_c + \mu_o \, ,
\end{equation}
where $V_c$ is the circular velocity of the galaxy, while $\mu_c$ (coefficient) and $\mu_o$ (offset) are free parameters. This is consistent with observations and in line with theoretical expectations that more massive galaxies have larger velocity offsets \citep{Yang2016,Verhamme2018}. We ignore the normalization constant since it will cancel out later when calculating the fraction of transmitted luminosity.

Since we assume emergent spectra for the observed models which are uncertain, we take an empirical approach by assuming idealized Gaussian profiles, but of course there are unmodelled processes including RT effects that induce anisotropic dust absorption, Doppler shifting, and line broadening. Despite these uncertainties, it is a well-defined optimisation problem to calibrate our model to observed LFs, and the resulting ratios of observed-to-intrinsic Ly$\alpha$ luminosities are relevant for understanding and interpreting observations of LAEs.

To calculate this transmission ratio, the line profile from each galaxy is multiplied by the frequency-dependent IGM transmission fraction of the central galaxy in its group, if known (see \ref{sec:Lya_methods} for details). Otherwise the line profiles from that group's galaxies are instead multiplied by the catalogue-median transmission curve. This is because we expect satellites to generally be fainter than the central but have similar cosmological scale IGM transmission with the exception of minor localized sightline and velocity offset effects. Therefore, we assume that satellites inherit transmission properties from the central halo and less-resolved galaxies inherit the median global transmission at each redshift. We integrate over frequency and then divide by the integral of the intrinsic profile:
\begin{equation}
  \mathcal{T}_\text{IGM} = \frac{\int \text{d}\nu\,J(\nu) e^{-\tau(\nu)}}{\int \text{d}\nu\,J(\nu)} \, .
\end{equation}
We then multiply $\mathcal{T}_\text{IGM}$ by the halo mass-dependent escape fraction from dust, $f_\text{esc}(M_\text{halo})$, which depends on the parameters of our dust model. This gives the final observed transmission fraction:
\begin{equation}
  f_\text{esc} \times \mathcal{T}_\text{IGM} = L_\alpha^\text{obs} / L_\alpha^\text{int} \, .
\end{equation}
From this, we calculate the observed Ly$\alpha$ luminosity, which gives us the observed LFs to calibrate to observations.

From our dust and spectral models, we thus have 5 parameters ($\log M_\text{halo}^\text{max}, \log M_\text{halo}^\text{max}, f_\text{esc}^\text{min}, \mu_\text{c}, \log \mu_\text{o}$) to optimize. Running a single model for all haloes is (relatively) computationally expensive, and thus standard minimisation techniques in such a high-dimensional parameter space can be costly. We employ a Gaussian Process Regression technique to build an emulator based on a suite of pre-performed simulations, allowing us to estimate the LF at various values of the free parameters in a rapid fashion. We use the \textsc{SWIFTEmulator} \citep{Kugel2022}, as this contains pre-written routines for emulating scaling relations like LFs.

We run the emulation at two redshifts, $z=5.5$ and $z=6.6$, generally following \citet{Morales2021}. We note that for the $z=5.5$ emulation, the observed data points we calibrated to were collected at $z=5.7$, however the reionization history (global \HI fraction, galaxy properties, etc.) between these two redshifts is very similar and the slight mismatch is unlikely to affect our results \citep[see][]{Kannan2022}. For the remainder of the paper, we refer to these observations as being at $z=5.5$. We also tried an emulation at $z=7$, but the available observations were not fully consistent with the LFs that \thesan predicts. Specifically, a Schechter function fit to the observed data points intersected the intrinsic LF at the faint-end.

We train the emulator with sufficient sampling within our five-dimensional parameter space, spread uniformly through the use of a Latin hypercube. The hypercube is generated with 300 realizations, from which we then reject parameter points that have $\log M_\text{halo}^\text{min} > \log M_\text{halo}^\text{max}$, an unphysical parameter combination, leaving 250 valid samplings. We list the ranges of parameters describing the critical volume of parameter space in Table~\ref{tab:params}, found after an initial exploration of the parameter space alongside physical considerations.

\begin{table}
  \centering
  \caption{Critical volume of parameter space for training the Gaussian Process emulator. We also require that $M_\text{halo}^\text{min} < M_\text{halo}^\text{max}$ for parameter viability. Note that the units for mass and $\mu_o$ are $\Msun$ and $\text{km\,s}^{-1}$, respectively.}
  \label{tab:params}
  \addtolength{\tabcolsep}{-1pt}
  \begin{tabular}{cccccc}
  \hline
  Range & $\log M_\text{halo}^\text{min}$ & $\log M_\text{halo}^\text{max}$ & $f_\text{esc}^\text{min}$ & $\mu_\text{c}$ & $\log \mu_\text{o}$ \\
  \hline
  $[\min,\max]$ & $[9.5,11.5]$ & $[10.5,12]$ & $[0.05,0.4]$ & $[0,2]$ & $[1,2]$ \\
  \hline
  \end{tabular}
  \addtolength{\tabcolsep}{1pt}
\end{table}

To train the emulator, we provide it with a vector specifying the location of the model in parameter coordinate space, i.e. $\vec{p} = (\log M_\text{halo}^\text{max}, \log M_\text{halo}^\text{max}, f_\text{esc}^\text{min}, \mu_\text{c}, \log \mu_\text{o})$, and the binned luminosity function $\phi$ constructed using an adaptive binning technique in $\log L_\alpha$ to control for Poisson error. We restrict the LF bin range to above $10^{41.5}\,\text{erg\,s}^{-1}$ to avoid extrapolating beyond observational sensitivities. Once trained, the emulator can make highly accurate predictions for $\phi(\vec{p}, L_\alpha)$ for any values of $\vec{p}$ and $L_\alpha$ covered by the hypercube and bins, respectively.

To find the best-fit parameters, we must compare our LFs to observed data. The Ly$\alpha$ LF shows the number density of LAEs in a given luminosity interval and is typically described using the \citet{Schechter1976} function:
\begin{equation} \label{eq:schechter}
  \phi(L)\text{d}(\log L) = \phi^\ast \left(\frac{L}{L^\ast}\right)^{\alpha+1} \exp\left(-\frac{L}{L^\ast}\right) \text{d}(\log L) \, ,
\end{equation}
where the redshift-dependent parameters are the normalization constant $\phi^\ast$, power-law slope for the faint end $\alpha$, and exponential cutoff scale for the bright end $L^\ast$. The rest-frame UVLF has been measured in detail out to $z \sim 10$ (Ly$\alpha$ LF out to $z \sim 7$) and is a powerful tool for studying the evolution of galaxy populations \citep[e.g.][]{Bouwens2015,Finkelstein2015,Oesch2015}.

Our observational data is collated from \citet{Ouchi2008}, \citet{Santos2016}, and \citet{Konno2018} for $z=5.5$, and from \citet{Ouchi2010}, \citet{Santos2016}, \citet{Konno2018}, and \citet{Taylor2020} for $z=6.6$. We fit Schechter functions to these data points using the {\tt curve\_fit} functionality available in the \textsc{SciPy} library. The best-fit parameters for our Schechter functions are summarized in Table~\ref{tab:schecter_catalog}. Note that as we are only using these Schechter fits to `smooth' out the data to simplify the interpretation of the fits, we do not fix any parameters in the Schechter fits. We then use the `penalty' functionality available in \textsc{SWIFTEmulator} to compute the average absolute offset between the Schechter fit and the predicted scaling relation. Here, we calculate
\begin{equation} \label{eq:penalty}
  I = \text{mean}_{L_\alpha} \left( \frac{|\phi_\text{fit}(L_\alpha) - \phi_\text{em}(\vec{p}, L_\alpha)|}{\epsilon}\right) \, ,
\end{equation}
with a fixed offset (also known as model discrepancy) of $\epsilon=0.4$, following the procedure in \citet{Rodrigues2017}, for all $L_\alpha$ values across the range. We calculate this penalty for both redshifts, and take the combined penalty to be their mean. Our best-fit model is the model with the lowest combined penalty, which is calculated using the minimisation routine present in \textsc{SciPy} along with the emulator.

\begin{table}
  \centering
  \caption{Best-fit parameters for capturing smooth representations of observed Ly$\alpha$ luminosity functions with Schechter functions (see equation~\ref{eq:schechter}). Note that the units for $\phi^\ast$ and $L^\ast$ are $\text{cMpc}^{-3}\,\text{dex}^{-1}$ and $\text{erg\,s}^{-1}$, respectively.}
  \label{tab:schecter_catalog}
  \begin{tabular}{cccc}
  \hline
  Redshift & $\log \phi^\ast$ & $\log L^\ast$ & $\alpha$ \\
  \hline
  $5.5$ & $-3.992$ & $43.45$ & $-2.534$ \\
  $6.6$ & $-4.123$ & $43.35$ & $-2.501$ \\
  \hline
  \end{tabular}
\end{table}

\begin{figure*}
  \centering
  \includegraphics[width=\textwidth]{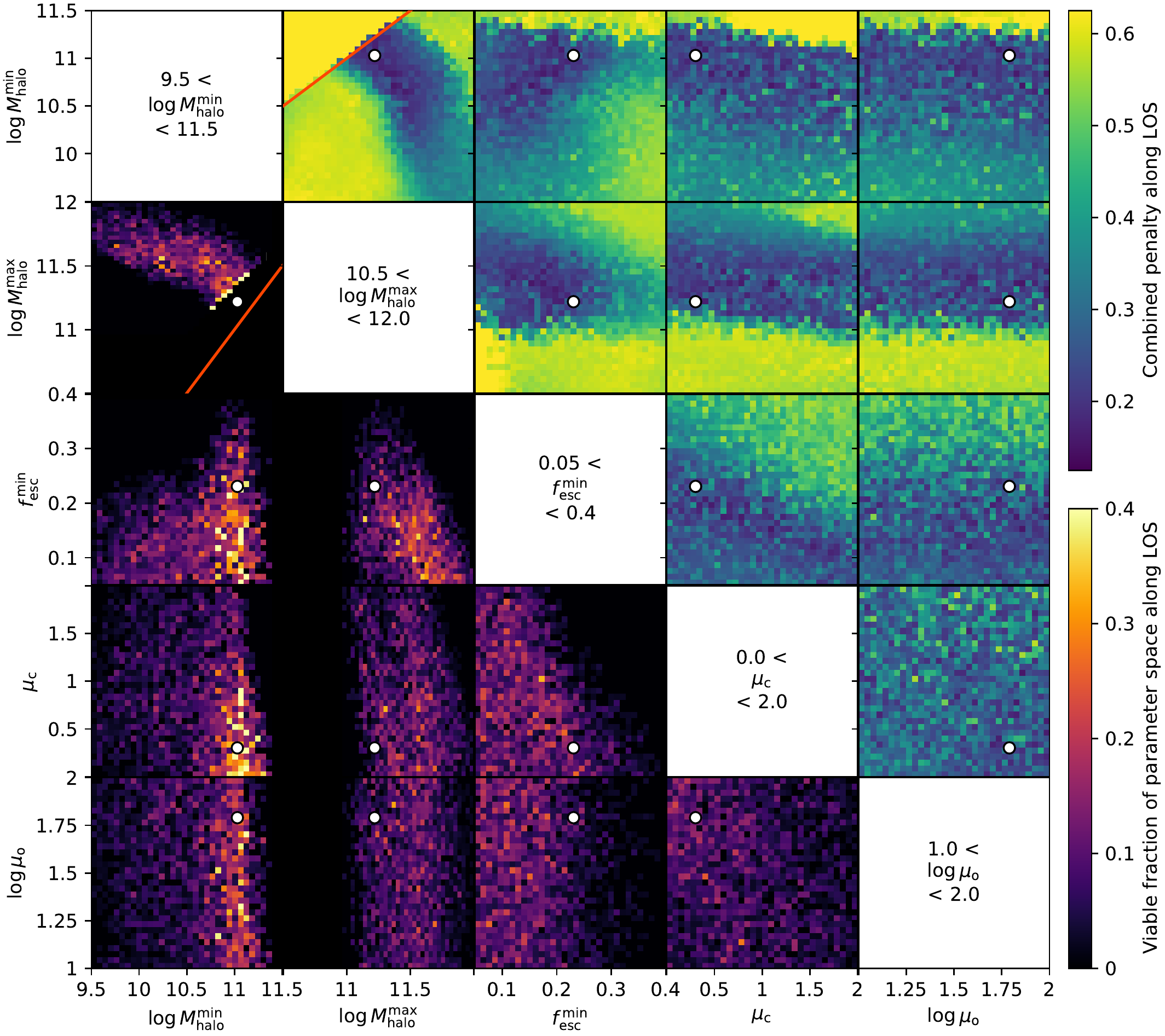}
  \caption{\textit{Upper right:} Combined penalty for $z=5.5$ and $z=6.6$ data, as a function of all marginalized parameter pairs. \textit{Lower left:} Volume fraction of parameter space along the line of sight spanned by implausibility $I < 0.6$. The red lines indicate the division where $M_\text{halo}^\text{min} > M_\text{halo}^\text{max}$ results in unphysical models. These views highlight the existence and prevalence of viable models, and place the best-fit model in the context of neighbouring penalty structures.}
  \label{fig:penalty_combined}
\end{figure*}

\begin{figure*}
  \centering
  \includegraphics[width=\textwidth]{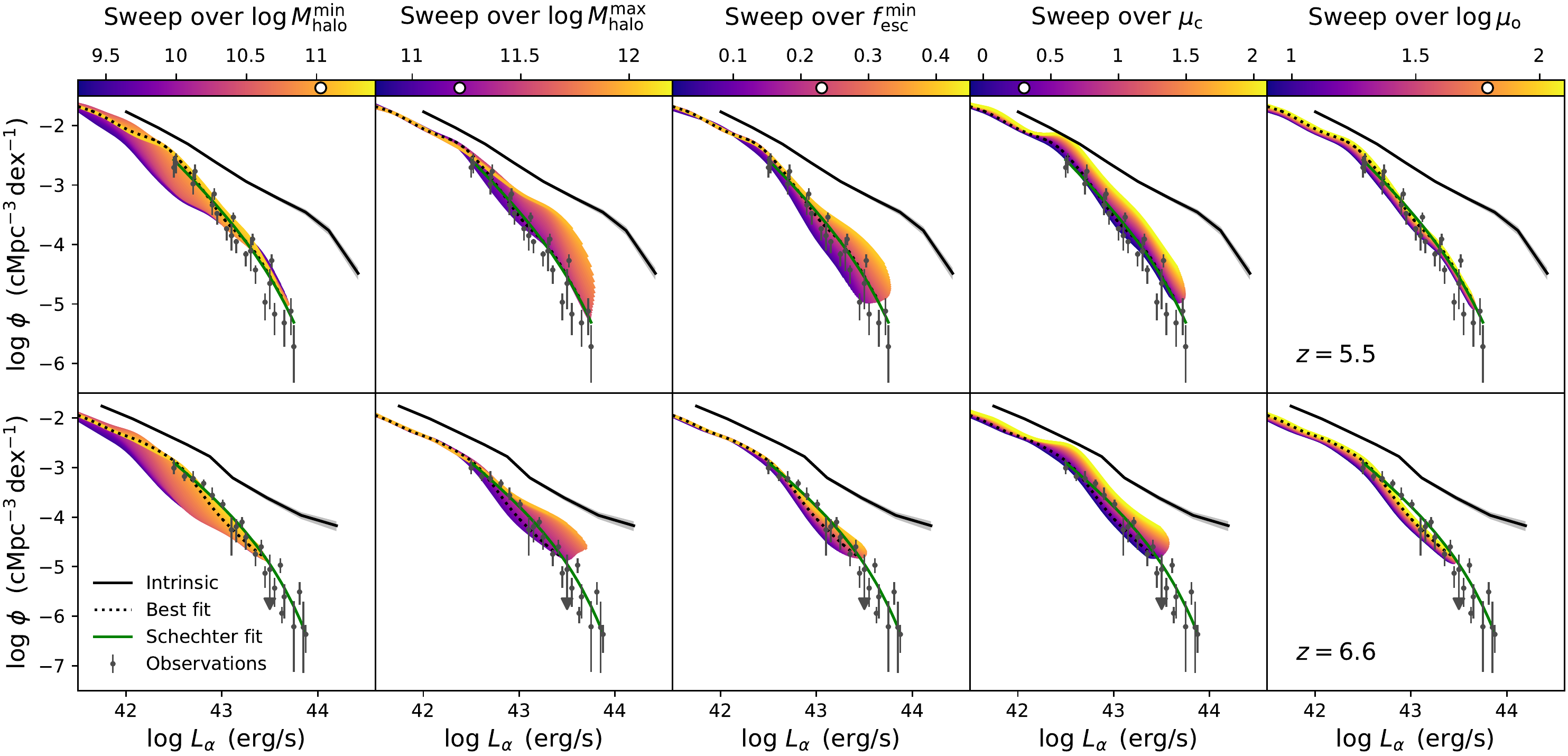}
  \caption{Sweep plots over all parameters, centred around the combined best fit for $z=5.5$ and $z=6.6$, with the best-fit model plotted as a dotted curve and also shown by white points in the colourbar axes. Observations are plotted in gray and the fitted Schechter function is in green. The intrinsic luminosity function is plotted in black. These trends confirm our physical intuition, and we also see that $\log \mu_\text{o}$ has less of an effect on the observed LF than the other parameters. We note that for sweeps over $\log M_\text{halo}^\text{min}$ and $\log M_\text{halo}^\text{max}$, we remove unphysical models with $M_\text{halo}^\text{min} > M_\text{halo}^\text{max}$.}
  \label{fig:fancysweep}
\end{figure*}

\subsection{Calibrated luminosity function results}
We now present the results of the calibration along with our interpretation of the model parameter space. In Fig.~\ref{fig:penalty_combined} we illustrate the penalty structure within the critical volume of parameter space, combining both $z=5.5$ and $z=6.6$ maximum absolute offsets between the fit and emulated values as calculated by equation~(\ref{eq:penalty}). For all parameter pairs, we show the combined penalty marginalized over the other model parameters, i.e. projected along the line of sight (LOS; upper right panels), and the volume fraction along the LOS spanned by implausibility defined by the penalty threshold $I < 0.4$ (lower left panels). Crucially, there is a small volume of parameter space that has an acceptable implausibility when matching both redshifts simultaneously, implying a tight constraint on many of our parameters. The best-fit model is shown as the white point within each panel, and we provide the specific parameter values in Table~\ref{tab:best_fit_model}, including for the individual redshifts and our fiducial combined model. We note that $\mu_\text{c}$, with a value of 0.3, is much lower than in other models, where it is generally around 1.5 (see Table~\ref{tab:catalog}). However, this may be compensated by having a dust model with a lower $f_\text{esc}^\text{min}$, or possibly a combination of adding a constant offset $\mu_\text{o}$ and different line widths. There is a clear indication in the panel of $f_\text{esc}^\text{min}$ vs. $\mu_\text{c}$ that in good models these parameters are inversely correlated, or at least in projection there is a wedge of viable combinations.

\begin{table}
  \centering
  \caption{Best-fit parameters as determined by the Gaussian Process Regression technique. Values are given for each redshift separately and for our fiducial model achieving the lowest combined mean penalty. Note that the units for mass and $\mu_o$ are $\Msun$ and $\text{km\,s}^{-1}$, respectively.}
  \label{tab:best_fit_model}
  \begin{tabular}{cccccc}
  \hline
  Redshift & $\log M_\text{halo}^\text{min}$ & $\log M_\text{halo}^\text{max}$ & $f_\text{esc}^\text{min}$ & $\mu_\text{c}$ & $\log \mu_\text{o}$ \\
  \hline
  5.5 & 10.36 & 11.49 & 0.1758 & 0.6657 & 1.949 \\
  6.6 & 11.02 & 11.34 & 0.2110 & 0.2750 & 1.851 \\
  Combined & 11.03 & 11.22 & 0.2307 & 0.3015 & 1.790 \\
  \hline
  \end{tabular}
\end{table}

In Table~\ref{tab:best_fit_model}, we also include parameter values for the best-fit model calculated for each redshift separately. The parameter space constraints for both redshifts are similar to Fig.~\ref{fig:penalty_combined} (not shown), and all values are within this similar range of good models. The individual best-fit parameters are also relatively similar, with the notable differences mainly indicating a slight redshift dependence on the dust or spectral model. Specifically, as expected the role of dust is more important at lower redshifts as $M_\text{halo}^\text{min}$ decreases; while the dust escape fraction stays more or less the same, the mass region affected by dust increases. For higher redshifts, the values of $\log M_\text{halo}^\text{min}$ and $\log M_\text{halo}^\text{max}$ are close together, making the escape fraction look close to a step function. In fact, Fig.~\ref{fig:penalty_combined} implies that this scenario for the escape fraction is the most reasonable, since the good models cluster near the red line where $\log M_\text{halo}^\text{min} = \log M_\text{halo}^\text{max}$. We note however that caveats to our dust model (detailed in section \ref{sec:implications}) mean that the shape of the $f_\text{esc}$ line should be taken with a grain of salt. Finally, we also see that $\mu_\text{c}$ increases from $z=6.6$ to $z=5.5$. However, the constraints on $\mu_\text{c}$ from Fig.~\ref{fig:penalty_combined} are fairly weak so this difference is likely due to the parameter degeneracies in fitting the LF.

\begin{figure}
  \centering
  \includegraphics[width=\columnwidth]{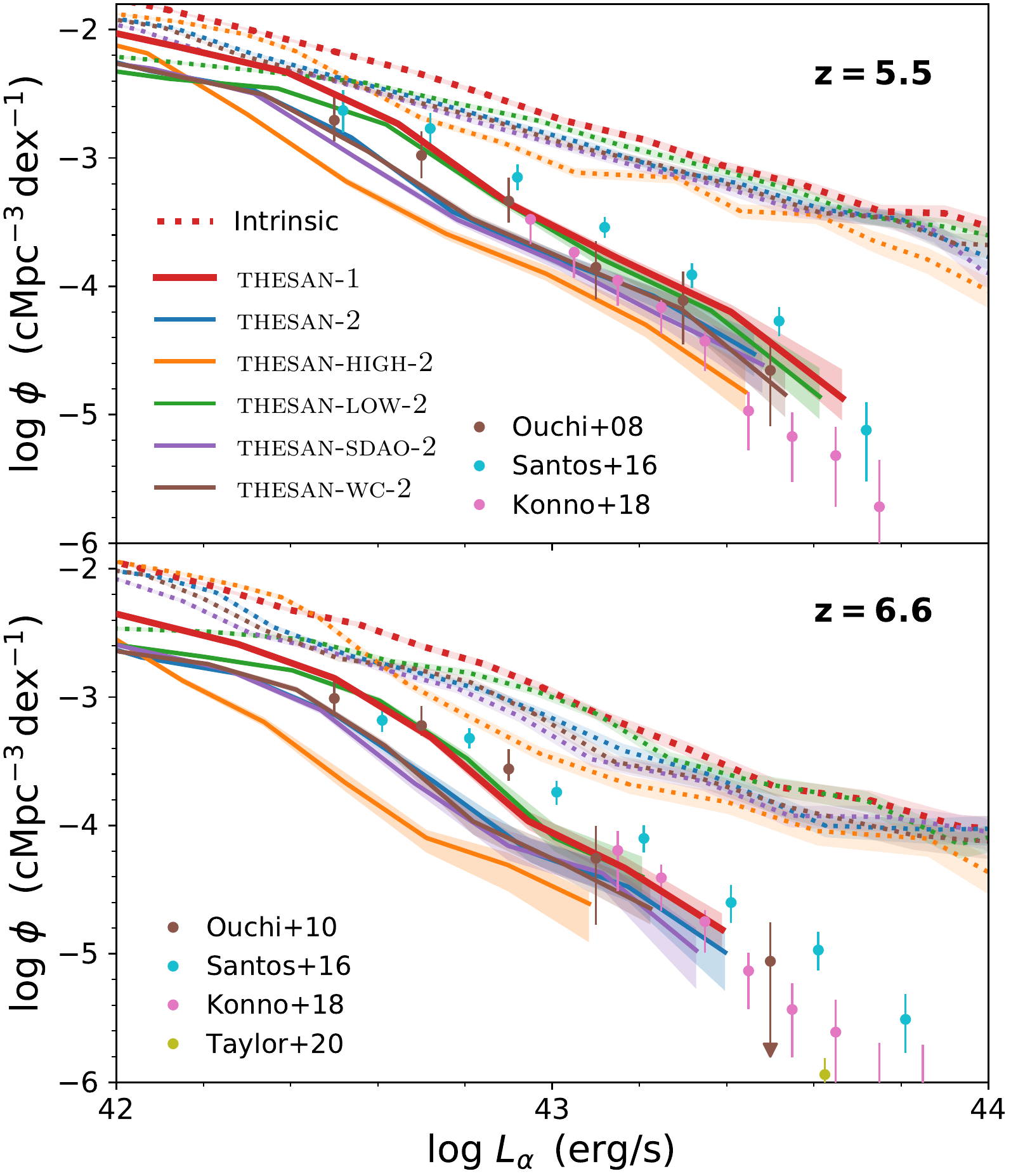}
  \caption{Comparison between the observed luminosity functions for all \thesan simulations after applying our best-fit model calibrated to \thesanone (thicker red line). Intrinsic luminosity functions are plotted as dotted lines. The ordering is preserved, indicating that physical differences affecting the intrinsic LFs are important in addition to differences in IGM transmission.}
  \label{fig:allsims}
\end{figure}

In Fig.~\ref{fig:fancysweep}, we plot sweeps over the range of each of the five parameters for both redshifts. Our best-fit model is plotted as a dotted line. The Schechter fits to the observations are shown in green, observations in gray, and the intrinsic LFs in black. The sweep for a particular parameter is performed by holding all other parameters constant at their best-fit values and predicting the LF for 50 values equally spaced over the specified parameter's range. This allows us to see how the specific parameter affects the LF. Specifically, we see that $\log \mu_\text{o}$ doesn't have much of an effect on the LF, while the other parameters provide much stronger constraints. This is supported by Fig.~\ref{fig:penalty_combined}, where the penalty plots for $\log \mu_\text{o}$ have a larger region of `good' models than the others.

Additionally, Fig.~\ref{fig:fancysweep} verifies that the parameters make physical sense and affect the LF in reasonable ways. For all the parameters, having a lower value corresponds to the LF moving to the left. This is expected, as both a lower $M_\text{halo}^\text{min}$ and $M_\text{halo}^\text{max}$ lead to dust absorption affecting lower mass, dimmer haloes, leading to more absorption overall, and a lower LF. A lower $f_\text{esc}^\text{min}$ also leads to more dust absorption. Finally, a smaller $\mu_\text{c}$ or $\mu_\text{o}$ means a smaller emergent spectral line offset, so the haloes are less luminous to begin with.

In Fig.~\ref{fig:allsims}, we apply our best-fit model to the entire suite of \thesan simulations and compare against observations. We find that all of the simulated LFs fit the observations reasonably well. Still, the simulations do not align with each other so the model calibration would need to be redone in each case to account for the differences between them. However, individual calibrations would effectively give a dust and spectral model that counteracts the physical differences distinguishing each simulation. Furthermore, these are mainly carried over from the differences in the intrinsic LFs (plotted as dotted lines) and IGM transmission, caused by factors explored in the previous section such as reionization history and bubble morphology. For example, we once again see \thesanlow and \thesanhigh bracketing the other simulations. To test the sensitivity of the calibration to the different physics of the simulations, however, we performed a calibration for \thesanwc data and found very similar constraints; more detailed results are given in Appendix~\ref{appendix:wc}.

\subsection{Implications for observed statistics}
\label{sec:implications}
Our spectral and dust model allows us to make broad predictions for observables that can be measured by current and future instruments, which we explore in this section. In Fig.~\ref{fig:visualization}, we show a visualization of the resolved recombination rate density along image sightlines calculated as $\int \alpha_\text{B} n_p n_e \text{d}V / \int \text{d}V$ similar to Equation~(\ref{eq:L_stars}) for a $95.5 \times 95.5 \times 2.3875\,\text{cMpc}^3$ subvolume of the box at $z=6.6$, along with points showing the intrinsic and observed Ly$\alpha$ luminosity of bright galaxies. We see that the background Ly$\alpha$ emission structure is highly correlated with both the intrinsic and observed luminosities, as expected, with galaxies of similar luminosities clustering together.

\begin{figure}
  \centering
  \includegraphics[width=\columnwidth]{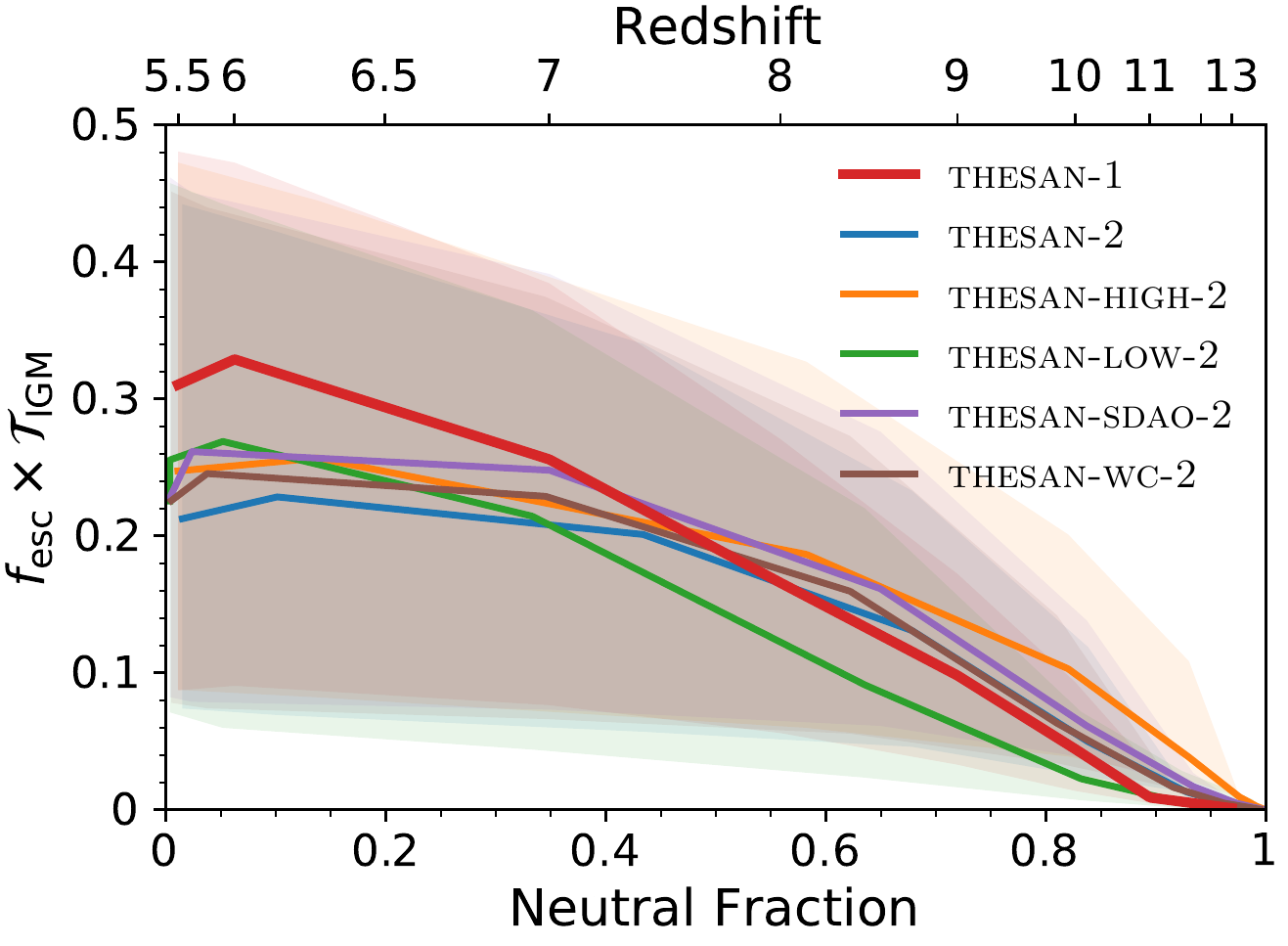}
  \caption{Ratio of observed-to-intrinsic Ly$\alpha$ luminosity ($f_\text{esc} \times \mathcal{T}_\text{IGM}$) for all \thesan simulations as functions of the global neutral fraction after applying our best-fit model. As in Fig.~\ref{fig:DW_xHI_200_400}, the sample is cut at $M_{1500} < -19$ (or brighter magnitudes depending on redshift) to avoid biases. The top redshift axis corresponds to the \thesanone reionization history.}
  \label{fig:X_z}
\end{figure}

In Fig.~\ref{fig:X_z}, we plot the ratio of observed-to-intrinsic Ly$\alpha$ luminosity including dust and IGM transmission effects ($L_\alpha^\text{obs} / L_\alpha^\text{int} = f_\text{esc} \times \mathcal{T}_\text{IGM}$) for all \thesan simulations as functions of the global neutral fraction. We applied the same best-fit dust and spectral model to all simulations and redshifts. As before, we cut the sample at intrinsic UV magnitudes of $M_{1500} < -19$ to avoid resolution biases across the runs, though similar to Figs.~\ref{fig:T_IGM_catalog}--\ref{fig:T_IGM_cdfetc} the equal weighting to all galaxies results in a bias towards lower masses. We also include redshift markings along the top axis corresponding to the \thesanone neutral fraction history. We do not apply an equivalent width lower limit to select for LAEs, but note that this ratio would be higher if we did, as this removes dust-obscured and transmission-suppressed galaxies.

The ordering of the simulations is slightly different from the plot of $\mathcal{T}_\text{IGM}$ in Fig.~\ref{fig:DW_xHI_200_400}. Here, \thesanone transmission is at the higher end of the simulations, while before it was lower. This is a reflection of differences in the intrinsic LFs and the stronger impact our model has on the other simulations for which it was not calibrated. As in Fig.~\ref{fig:DW_xHI_200_400}, \thesanlow and \thesanhigh once again bracket the medium resolution simulations early on due to reionization history and morphology, with \thesanlow below the others and \thesanhigh above, but all simulations are roughly equivalent below $z \approx 7$. Another similarity is the more or less linear progression upwards as neutral fraction decreases, though of course the value does not increase to unity. Instead, \thesanone peaks around $30^{+10}_{-20}\%$ showing significant sightline-to-sightline and galaxy-to-galaxy variations, as summarized in Table~\ref{tab:X}.

\begin{table}
  \centering
  \caption{Ratios of observed-to-intrinsic Ly$\alpha$ luminosity ($f_\text{esc} \times \mathcal{T}_\text{IGM}$) and covering fractions defined as the fraction of sightlines for each galaxy with ratios below 10 per cent $P(<0.1)$ for the \thesanone simulation tabulated as functions of redshift (as percentages). The summary statistics are calculated including all central galaxies with UV brightness $M_{1500} < -19$ and are given in per cent units with median and asymmetric 1$\sigma$ confidence regions.}
  \label{tab:X}
  \addtolength{\tabcolsep}{-2pt}
  \renewcommand{\arraystretch}{1.1}
  \begin{tabular}{ccccccc}
  \hline
  Quantity & $z=6$ & $z=7$ & $z=8$ & $z=9$ & $z=10$ \\
  \hline
  $f_\text{esc} \times \mathcal{T}_\text{IGM}$ & ${32.9}^{+14.4}_{-23.8}$ & ${25.6}^{+12.9}_{-17.9}$ & ${16.6}^{+10.5}_{-11.0}$ & ${9.8}^{+7.4}_{-6.5}$ & ${4.4}^{+4.9}_{-3.1}$ & \vspace{.1cm} \\
  $P(<0.1)$ & ${6.8}^{+49.9}_{-2.5}$ & ${10.4}^{+7.2}_{-3.0}$ & ${20.1}^{+9.5}_{-5.2}$ & ${51.0}^{+13.7}_{-15.6}$ & ${89.1}^{+6.8}_{-10.7}$ \\
  \hline
  \end{tabular}
  \addtolength{\tabcolsep}{2pt}
  \renewcommand{\arraystretch}{0.9090909090909090909}
\end{table}

\begin{figure}
  \centering
  \includegraphics[width=\columnwidth]{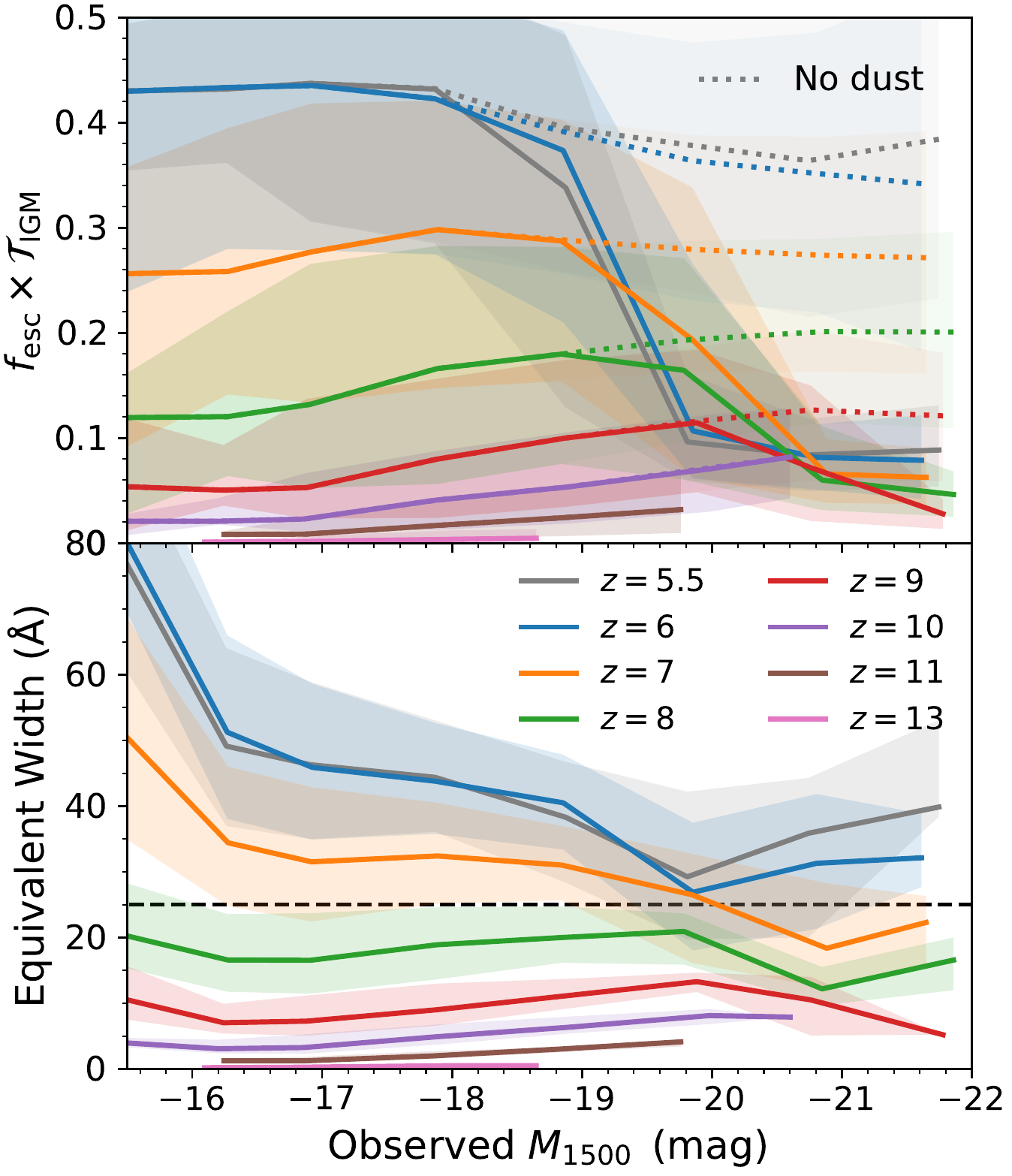}
  \caption{\textit{Upper panel:} Observed-to-intrinsic Ly$\alpha$ luminosity ($f_\text{esc} \times \mathcal{T}_\text{IGM}$) and transmission without dust ($\mathcal{T}_\text{IGM}$; dotted curves) as functions of observed $M_{1500}$ for \thesanone at several redshifts. \textit{Lower panel:} Observed rest-frame EW as a function of observed $M_{1500}$. As described in the text, we limit the continuum escape fractions by our best-fit model Ly$\alpha$ escape fractions to avoid unrealistic escape scenarios. Despite the lack of non-LAE massive galaxies, the redshift evolution around $M_{1500} \sim -19$ roughly agrees with our expectation for the declining LAE fraction in the pre-reionization epoch.}
  \label{fig:EW}
\end{figure}

In the top panel of Fig.~\ref{fig:EW}, we show the observed-to-intrinsic luminosity ($f_\text{esc} \times \mathcal{T}_\text{IGM}$) and transmission without dust ($\mathcal{T}_\text{IGM}$; dotted curves)
as functions of the observed UV magnitude for \thesanone at several redshifts. Including dust absorption causes the value to begin decreasing around a magnitude of $-18$ to $-19$, as expected due to the absorption fraction increasing with magnitude. Without dust absorption our model exhibits a relatively flat dependence on UV magnitude, which is a bit counter intuitive because the IGM transmission bands show an increase with $M_{1500}$ and strong infall velocity signatures. However, when considering the emergent spectral profile this seems to be erased, likely because brighter galaxies have larger velocity offsets.

While there is a correspondence between the observed-to-intrinsic luminosity and the fraction of galaxies that are LAEs, after completing our analysis we discovered the idealized dust model is incompatible with corrections typically used for $M_\text{UV}$ \citep{Gnedin2014,KannanMTNG}, so we do not predict realistic rest-frame equivalent widths (EWs) calculated as $\text{EW} = L_\alpha^\text{obs} / L_{\lambda,\text{cont}}^\text{obs}$. Therefore, the predicted observed-to-intrinsic luminosity ratios should be viewed as statistically correct; i.e. they reproduce observed LFs, but the manner in which this is achieved yields too high of EWs for the most massive galaxies. This is because the Ly$\alpha$ escape fraction cannot be lower than $f_\text{esc}^\text{min}$ while the UV escape fraction can be arbitrarily low calculated as $\exp(-\tau)$. The degeneracy between dust absorption and IGM transmission effectively means that our massive galaxies have too high $f_\text{esc}$ and too low $\mathcal{T}_\text{IGM}$. However, we emphasize that the general $f_\text{esc} \times \mathcal{T}_\text{IGM}$ predictions remain trustworthy, e.g. UV bright galaxies have significant dust absorption while fainter objects ($M_{1500} \gtrsim -19$) are more direct probes of the IGM evolution. In any case, this motivates a future analysis with self-consistent Ly$\alpha$ and continuum dust corrections.

To qualitatively understand the evolution of the LAE fraction \citep{Stark2011,Schenker2014,Mason2019}, in the bottom panel of Fig.~\ref{fig:EW}, we also plot a possible prediction of the rest-frame EW as a function of observed $M_{1500}$. We calculate the observed continuum escape fractions following the \citet{Gnedin2014} correction and our best-fit model for the observed Ly$\alpha$ luminosities. The line continuum value at $1216\,\text{\AA}$ is extrapolated from the dust-corrected values at $1500$ and $2500\,\text{\AA}$, i.e. assuming power-law UV slopes \citep[see section 3.2 of][]{Smith2022}. However, realizing that the dust models are incompatible we also limit the UV dust correction escape fraction for each galaxy so it does not dip below our Ly$\alpha$ escape fraction, as otherwise it would not be physical. For reference, we plot a horizontal dotted grey line at $25\,\text{\AA}$ to mark the minimum threshold to be classified as an LAE. We see that starting at $z \lesssim 7$ fainter galaxies at observed UV magnitudes $M_{1500} \gtrsim -20$ largely fall above this line. This is exactly the LAE fraction evolution that we expect. However, the lack of Lyman break galaxies, i.e. with low-EWs and therefore not LAEs, is not realistic. This is again due to the escape fraction for the continuum correction around the line going down to zero, while our dust escape fraction for Ly$\alpha$ plateaus at $f_\text{esc}^\text{min}$ after $M_\text{halo}^\text{max}$, so our predicted EWs with escape fraction limiting are now well-behaved but are still generally too high at the bright end.

This problem would likely be solved with another dust model where the dust escape fractions for Ly$\alpha$ photons are required to be less than or equal to those of continuum photons. This would naturally reduce the EWs of the most massive galaxies $M_\text{halo} \gtrsim 10^{11.5}\,\Msun$ to below LAE thresholds $25\,\text{\AA}$ at all redshifts. We leave this to a future study and include our current model as a first exploration of the emulated calibration technique that succeeds in matching observed Ly$\alpha$ LFs but requires improvements to predict realistic EWs.

\section{Conclusions}
\label{sec:conclusions}
In this paper, we have constructed catalogues for Ly$\alpha$ emission and transmission for the entire suite of \thesan simulations. The \thesan simulations have been shown to provide a unique framework for statistically studying Ly$\alpha$ properties throughout the EoR \citep{Smith2022}, and we have extended the analysis to include comparisons between the medium resolution simulations, each with different reionization histories and bubble morphologies. This allowed us to explore the effects of spatial resolution, alternative dark matter models, halo mass-dependent escape fractions, and numerical convergence on the frequency-dependent IGM transmission throughout the EoR ($5.5 \leq z \leq 13$). It is unclear how to interpret Monte Carlo Ly$\alpha$ radiative transfer calculations in the context of the galaxy formation model used in \thesan. Therefore, we chose to employ an empirical model for the dust escape fractions and emergent spectral profiles to act as a bridge between the much more robust intrinsic emission and IGM transmission predictions. We then used a Gaussian Process Regression technique to calibrate our model to optimally reproduce the observed Ly$\alpha$ luminosity function. The dust and spectral models have degenerate absorption properties, which complicates our ability to uniquely determine the precise physical origin of the large difference between intrinsic and observed LFs. However, we were still able to find a small volume of parameter space with good models that fit observations, which demonstrates the power of this type of empirical modelling. We summarize our main conclusions as follows:
\begin{enumerate}
  \item In all simulation variations, transmission redward of line centre increases with decreasing redshift while transmission blueward is highly suppressed until around $z \approx 6$, and remains low throughout the tail-end of reionization, as expected (see Fig.~\ref{fig:T_IGM_catalog}).
  \item Transmission around the red peak is also highly sensitive to frequency. For a waveband centred at $\Delta v = 200\,\text{km\,s}^{-1}$, transmission is more suppressed and has more variation than at $\Delta v = 400\,\text{km\,s}^{-1}$, since it is more affected by the local environment and infalling gas (see Fig.~\ref{fig:DW_200_400}).
  \item \thesanlow has the earliest reionization history due to the dominant role of low-mass galaxies contributing to reionization. On the other hand, \thesanhigh is delayed, as high-mass galaxies are much rarer at early times. This leads to differences in the transmission curves: \thesanlow transmission is boosted while \thesanhigh transmission is more suppressed for all redshifts.
  \item When correcting for reionization history by looking at the global neutral fraction instead, the red peak transmission increases nearly linearly with decreasing neutral fraction. The difference between simulations also highlights the importance of bubble morphology on transmission, with \thesanhigh having an advantage compared to \thesanlow at the same neutral fraction (see Fig.~\ref{fig:DW_xHI_200_400}).
  \item The variation across sightlines for a single galaxy is greater than the variation globally across all galaxies, mainly because a galaxy's environment can look significantly different in different directions (see Fig.~\ref{fig:T_IGM_cdfetc}).
  \item Red peak transmission increases with UV magnitude $M_{1500}$ at $z \gtrsim 6$, except near line centre, when there is a downturn for bright galaxies because of infalling gas. This behaviour is mirrored in the corresponding covering fractions defined as the fraction of sightlines around each galaxy with low transmission $P(\mathcal{T}_\text{IGM} < 0.2)$ (see Figs.~\ref{fig:T_IGM_M1500} and \ref{fig:covering_fractions}).
  \item Previous empirical models do not account for dust absorption, however we find doing so is essential in realistic comparisons to observed luminosity functions, as this becomes the dominant mechanism for removing Ly$\alpha$ photons in massive galaxies at $z \lesssim 7$ (see Figs.~\ref{fig:nodustcomparison} and \ref{fig:EW}).
  \item In our idealized dust and spectral model, the minimum dust escape fraction ($f_\text{esc}^\text{min}$) and spectral profile offset from line centre ($\mu_\text{c}$) are inversely correlated in the space for good models, as both cause a lower observed LF. Our framework also allows us to understand model degeneracies more generally (see Figs.~\ref{fig:penalty_combined} and \ref{fig:fancysweep}).
  \item Intrinsic luminosity functions and IGM transmission curves vary across different \thesan simulations, so parameter recalibration of the ratio of observed-to-intrinsic Ly$\alpha$ luminosity ($f_\text{esc} \times \mathcal{T}_\text{IGM}$) is recommended for different simulations. We found our best-fit model works well for \thesanone and is reasonable for the others, except in the case of \thesanhigh (see Figs.~\ref{fig:allsims} and \ref{fig:X_z}).
  \item The dust escape fractions for Ly$\alpha$ and stellar continuum photons must be modelled self-consistently. Otherwise, it is possible to have $f_\text{esc}^{\text{Ly}\alpha} > f_\text{esc}^\text{cont}$ and the model overpredicts the equivalent widths of high mass galaxies. However, due to dust--spectral degeneracies, realistic predictions for observable LAE statistics beyond LFs are still possible (see Fig.~\ref{fig:EW}).
\end{enumerate}

In the future we plan to design and calibrate a dust model in which Ly$\alpha$ escape fractions are limited by continuum ones. Allowing $f_\text{esc}$ to approach zero asymptotically for high mass galaxies will likely lead to more reasonable EW predictions. Such an improvement will expand the utility of our catalogues for further applications, including understanding the clustering of LAEs, 21\,cm cross-correlations, connections to quasar absorption troughs, and other Ly$\alpha$-centric observational probes of reionization. Overall, constrained empirical modelling represents a promising avenue to elucidate LAE science, especially when combined with state-of-the-art large-volume galaxy formation and reionization simulations with realistic Ly$\alpha$ emission and transmission.

\section*{Acknowledgements}
We thank the referee for constructive comments and suggestions which have improved the quality of this work.
We thank Luz \'Angela Garcia, Charlotte Mason, Alexa Morales, and Lucia Perez for insightful discussions related to this work. CX acknowledges support from the MIT Undergraduate Research Opportunities Program (UROP), specifically the Paul E. Gray (1954) UROP Fund. AS acknowledges support for Program number \textit{HST}-HF2-51421.001-A provided by NASA through a grant from the Space Telescope Science Institute, which is operated by the Association of Universities for Research in Astronomy, incorporated, under NASA contract NAS5-26555. MV acknowledges support through NASA ATP grants 16-ATP16-0167, 19-ATP19-0019, 19-ATP19-0020, 19-ATP19-0167, and NSF grants AST-1814053, AST-1814259, AST-1909831 and AST-2007355. The authors gratefully acknowledge the Gauss Centre for Supercomputing e.V. (\url{www.gauss-centre.eu}) for funding this project by providing computing time on the GCS Supercomputer SuperMUC-NG at Leibniz Supercomputing Centre (\url{www.lrz.de}). Additional computing resources were provided by the MIT Engaging cluster. We are thankful to the community developing and maintaining software packages extensively used in our work, namely: \texttt{matplotlib} \citep{matplotlib}, \texttt{numpy} \citep{numpy} and \texttt{scipy} \citep{scipy}.

\section*{Data Availability}
All \thesan simulation data will be made publicly available in the near future, including Ly$\alpha$ catalogues. Data will be distributed via \url{www.thesan-project.com}. Before the public data release, data underlying this article will be shared on reasonable request to the corresponding author(s).


\bibliographystyle{mnras}
\bibliography{biblio}



\appendix

\section{Resolution comparison}
\label{appendix:resolution}

\begin{figure}
  \centering
  \includegraphics[width=\columnwidth]{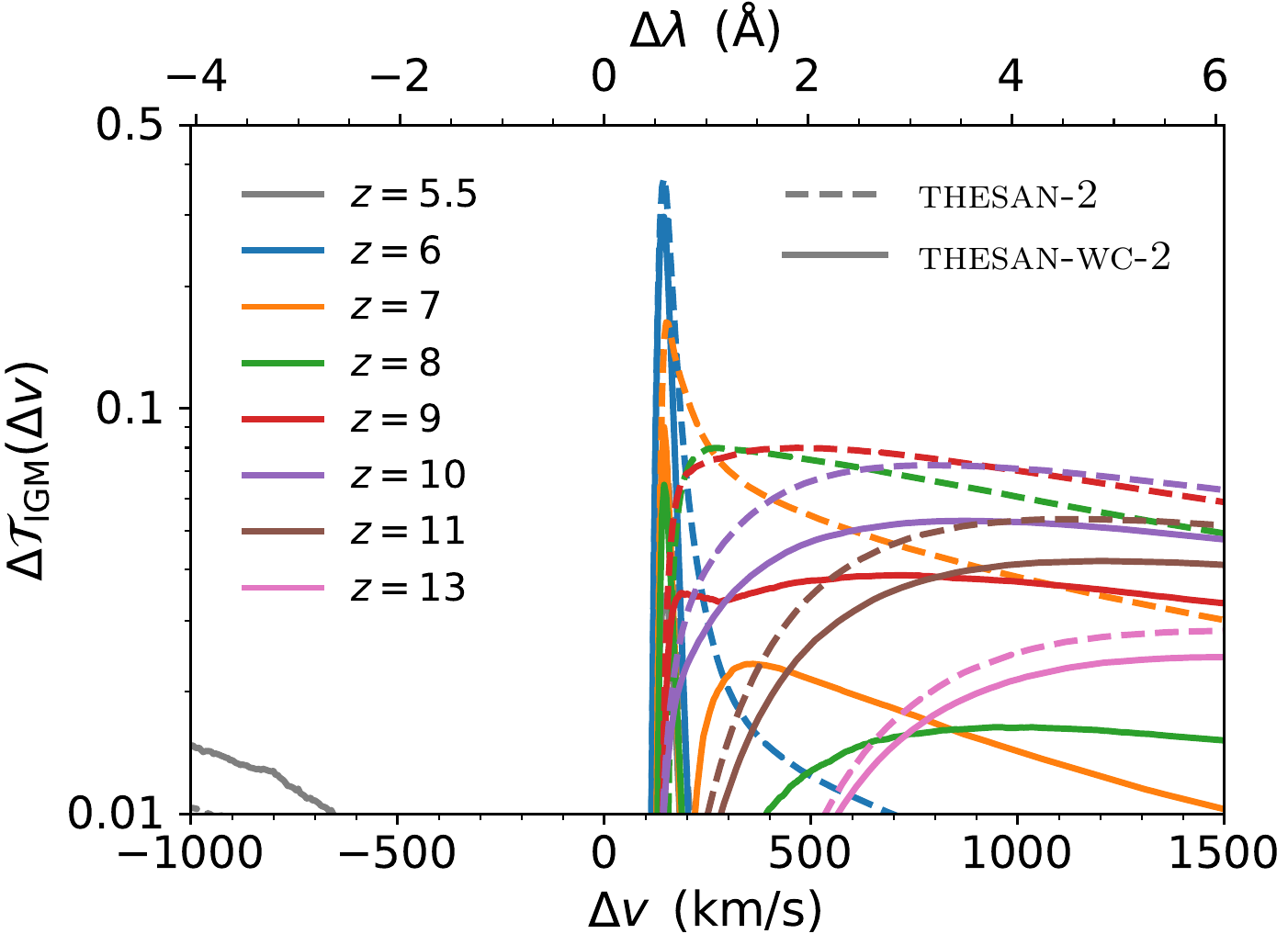}
  \caption{The absolute difference $\Delta \mathcal{T}_\text{IGM} \equiv |\mathcal{T}_\text{IGM,1} - \mathcal{T}_\text{IGM,2}|$ between \thesanone and \thesantwo (dashed lines) and \thesanwc (solid lines) simulations, shown for median statistics. This spatial resolution convergence comparison reveals a strong feature corresponding to the location of the step function transition and overall agreement at the level of $\Delta \mathcal{T}_\text{IGM} \lesssim 0.05$ for the weak convergence run elsewhere.}
  \label{fig:T_IGM_resolution}
\end{figure}

Throughout this paper we have compared various results from \thesanone and the medium resolution simulations. In this section we directly examine where the main differences are in the frequency-dependent IGM transmission curves. Due to chaotic variations in galaxy formation modelling, we do not expect identical results at the level of individual haloes even for simulations run with the same random number generator seeds. However, it is reasonable to compare the median (or mean) transmission curves between \thesanone and \thesantwo, which has a factor of eight lower mass resolution but is otherwise the same. In Fig.~\ref{fig:T_IGM_resolution} we show the absolute difference $\Delta \mathcal{T}_\text{IGM} \equiv |\mathcal{T}_\text{IGM,1} - \mathcal{T}_\text{IGM,2}|$ for the \thesantwo and \thesanwc simulations. These both show a strong feature at $\Delta v \approx 100$--$200\,\text{km\,s}^{-1}$ resulting from the slightly different locations of the transition from blue suppression to red transmission, which occurs closer to line center for \thesanone as a result of resolving additional dense substructure for larger covering fractions of inflowing \HI gas. There are also differences in the red damping-wing transmission at $\Delta v \gtrsim 300\,\text{km\,s}^{-1}$, which can be explained by differences in the reionization history and bubble morphology. The weak convergence run partially accounts for these differences by increasing the birth cloud escape fraction from $0.37$ to $0.43$, and this is reflected in the comparison in this region with $\Delta \mathcal{T}_\text{IGM} < 0.08$ for \thesantwo and much lower for \thesanwc in some redshift cases. Overall, the impact of resolution in the IGM (outside $R_\text{vir}$) is minor.

\section{\thesanwc calibration}
\label{appendix:wc}

\begin{figure*}
  \centering
  \includegraphics[width=\textwidth]{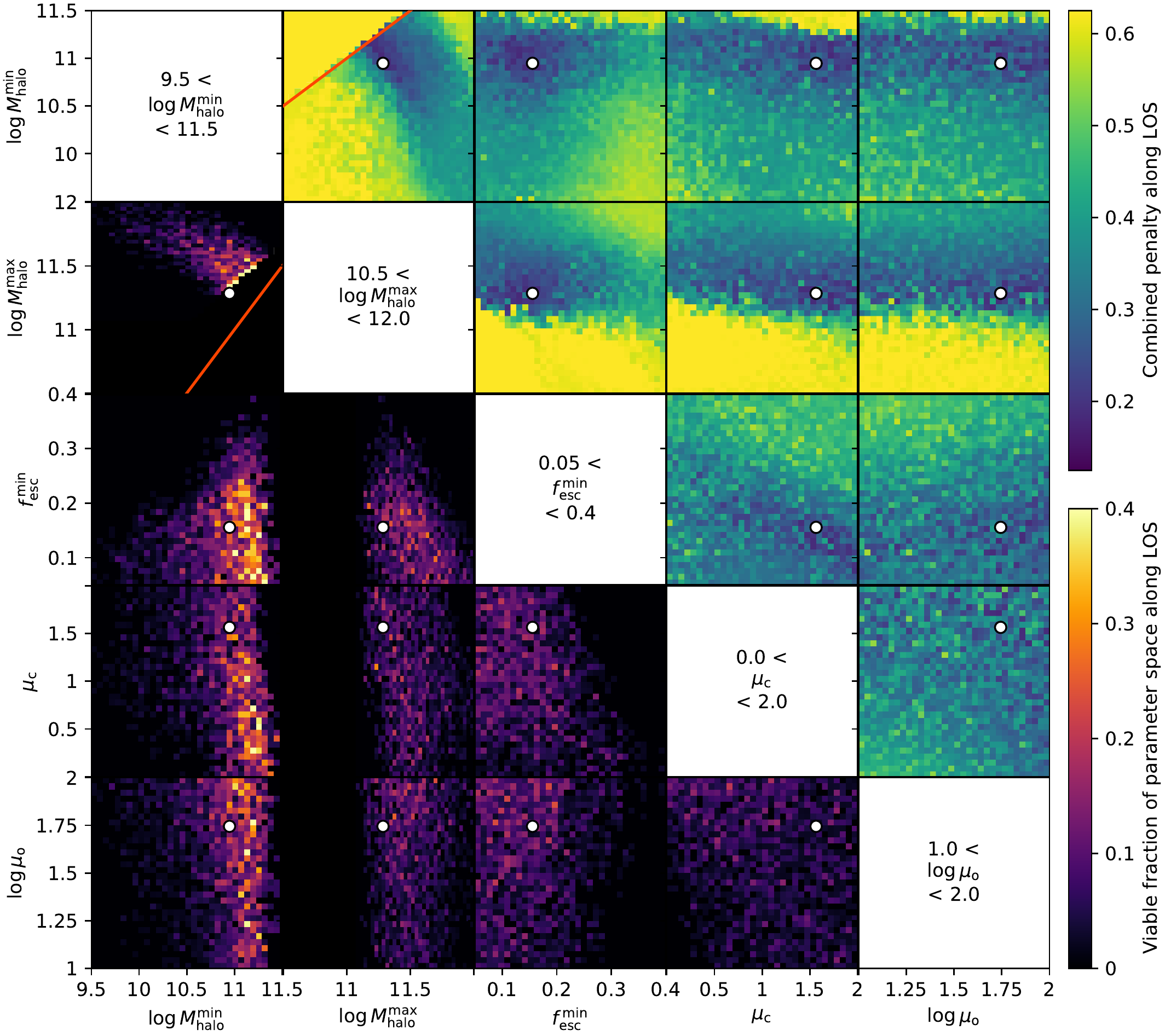}
  \caption{Counterpart to Fig.~\ref{fig:penalty_combined} for the \thesanwc calibration. \textit{Upper right:} Combined penalty for $z=5.5$ and $z=6.6$ data. \textit{Lower left:} Volume fraction of parameter space along the line of sight spanned by implausibility $I < 0.6$. The red lines indicate the division where $M_\text{halo}^\text{min} > M_\text{halo}^\text{max}$ results in unphysical models. The penalty structure for \thesanwc shows the same shape as the one for \thesanone, implying that the resolution did not significantly affect the calibration.}
  \label{fig:wc_penalty}
\end{figure*}

\begin{figure}
  \centering
  \includegraphics[width=\columnwidth]{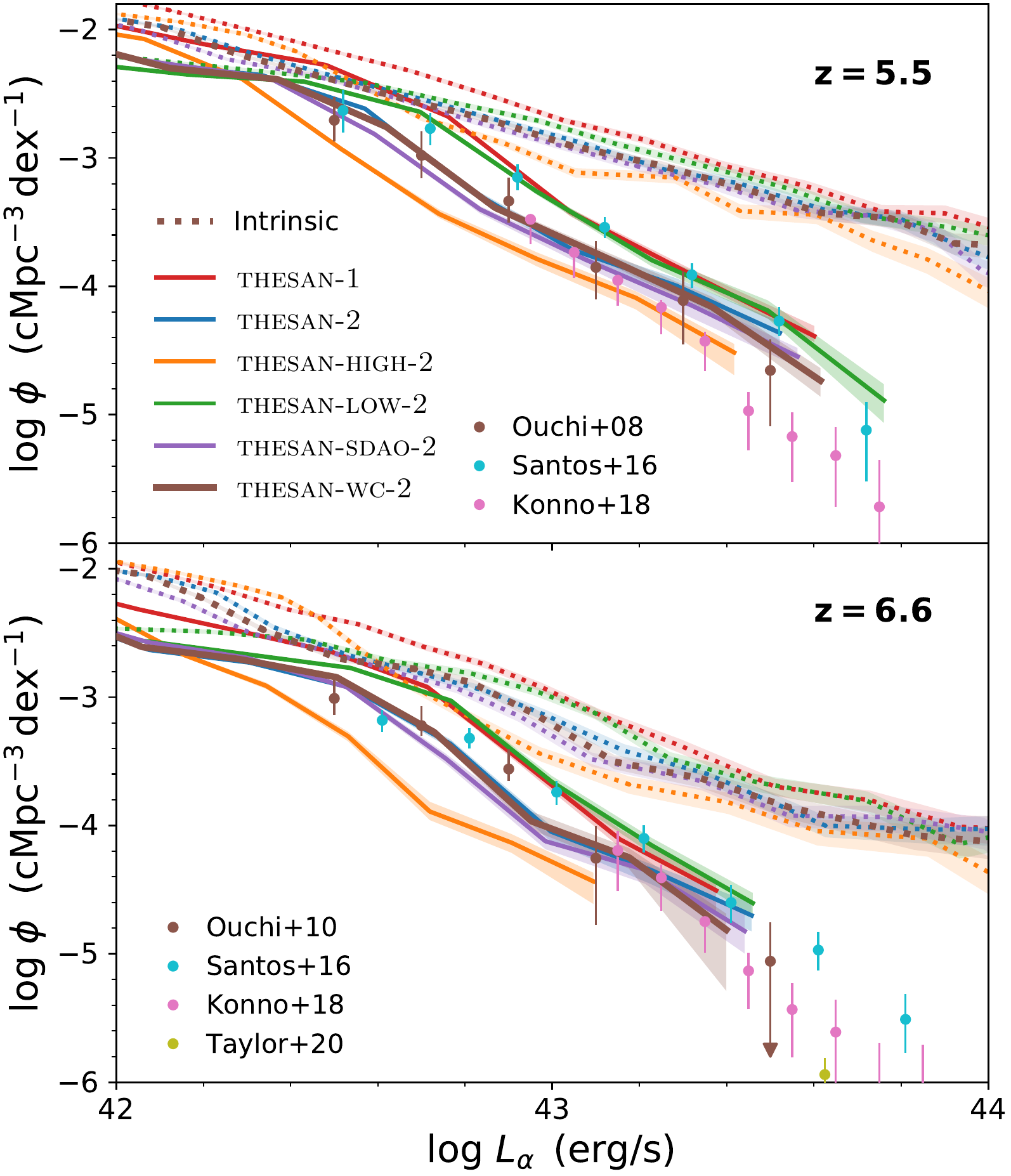}
  \caption{Comparison between the observed luminosity functions for all \thesan simulations after applying our best-fit model calibrated to \thesanwc. Intrinsic luminosity functions are plotted as dotted lines. Much like Fig.~\ref{fig:allsims}, the ordering is preserved and all LFs match observations reasonably well.}
  \label{fig:wc_LFallsims}
\end{figure}

Our dust and spectral model calibration in the main body of the paper was only done for \thesanone. Here we present the results of the same calibration routine applied to \thesanwc. The best-fit model, given in Table~\ref{tab:wc_best_fit_model}, has a higher $\mu_\text{c}$ and lower $f_\text{esc}^\text{min}$ than was found for \thesanone. However, the penalty structure in Fig.~\ref{fig:wc_penalty} is extremely similar to Fig.~\ref{fig:penalty_combined}, with the same negative correlation between $\mu_\text{c}$ and $f_\text{esc}^\text{min}$. It is therefore more likely that differences in the best-fit model are mainly due to parameter degeneracies rather than statistically significant physical differences. We also applied this best-fit model to all simulations at $z=5.5$ and $6.6$, the results of which are shown in Fig.~\ref{fig:wc_LFallsims}. Again, all simulated LFs match observations, with differences carrying over from the intrinsic LFs. Overall, the simulation resolution did not have a significant effect on calibration results other than correcting slightly for physical differences.

\begin{table}
  \centering
  \caption{Best-fit parameters for \thesanwc. Note that the units for mass and $\mu_o$ are $\Msun$ and $\text{km\,s}^{-1}$, respectively.}
  \label{tab:wc_best_fit_model}
  \begin{tabular}{cccccc}
  \hline
  Redshift & $\log M_\text{halo}^\text{min}$ & $\log M_\text{halo}^\text{max}$ & $f_\text{esc}^\text{min}$ & $\mu_\text{c}$ & $\log \mu_\text{o}$ \\
  \hline
  5.5 & 11.07 & 11.24 & 0.1509 & 1.6986 & 1.663 \\
  6.6 & 11.18 & 11.33 & 0.0762 & 1.5284 & 1.316 \\
  Combined & 10.95 & 11.28 & 0.1555 & 1.5648 & 1.744 \\
  \hline
  \end{tabular}
\end{table}


\bsp	
\label{lastpage}
\end{document}